\documentclass[10pt, a4paper]{iopart-per}
\LogoHeight{1.8cm}%

\usepackage{iopams} 
\usepackage{subfig,graphicx}
\usepackage{rotating}
\usepackage[utf8]{inputenc}
\begin{document}

\title[On the triggering of the Ultimate Regime of convection]{On the triggering of the Ultimate Regime of convection}

\author{P.-E. Roche, F. Gauthier, R. Kaiser and J. Salort}

\address{Institut N\'eel, CNRS / UJF - BP 166, F-38042 Grenoble cedex 9, France}
\begin{abstract}
Rayleigh-Bénard cells are one of the simplest systems to explore the laws of natural convection in the highly  turbulent limit. However, at very high Rayleigh numbers ($Ra \gtrsim 10^{12}$) and for Prandtl numbers of order one, experiments fall into two categories : some evidence a steep enhancement of the heat transfer while others do not. The origin of this apparent disagreement is presently unexplained. This puzzling situation motivated a systematic study of the triggering of the regime with an enhanced heat transfer, originally named the ``Ultimate Regime'' of convection. High accuracy heat transfer measurements have been conducted in convection cells with various aspect ratios and different specificities, such as altered boundary conditions or obstacles inserted in the flow. The two control parameters, the Rayleigh and Prandtl numbers have been varied independently to disentangle their relative influence.  Among other results, it is found that i) most experiments reaching very high $Ra$ are not in disagreement if small differences in Prandtl numbers are taken into account, ii) the transition is not directly triggered by the large scale circulation present in the cell, iii) the sidewall of the cell have a significant influence on the transition. The characteristics of this Ultimate regime are summarized and compared with R. Kraichnan prediction for the asymptotic regime of convection.

\end{abstract}

\maketitle

\section{Introduction: an elusive regime }

\subsection{An historical perspective}

In 1996, an abrupt enhancement of the heat transport efficiency ($Nu$) was reported in a convection cell driven at very high Rayleigh numbers ($Ra$) \cite{Chavanne1996,Chavanne1997} (the definitions of $Ra$ and $Nu$ are recalled later).
This observation was understood as the signature of a new regime of convection, named the ``Ultimate'' regime, and was interpreted following a prediction by R. Kraichnan \cite{Kraichnan1962}.
However this observation was in apparent contradiction with some earlier $Nu(Ra)$ measurements which did not evidence any new regime in nearly similar conditions  \cite{WuTHESE}. 
This situation ignited a controversy which has grown up over the years, as additional observations seemed to confirm both the transiting dataset from Grenoble and the non-transiting one from Chicago.
Today, this issue is often considered as one of the most important open problem in convection and is driving experimental and numerical efforts worldwide. The ability to extrapolate laboratory results to environmental flows, for example, is strongly impaired by our lack of understanding of turbulent convection at very high Rayleigh numbers \cite{SommeriaPREFACE}.

\begin{figure}
\center
\includegraphics[width=.8\columnwidth]{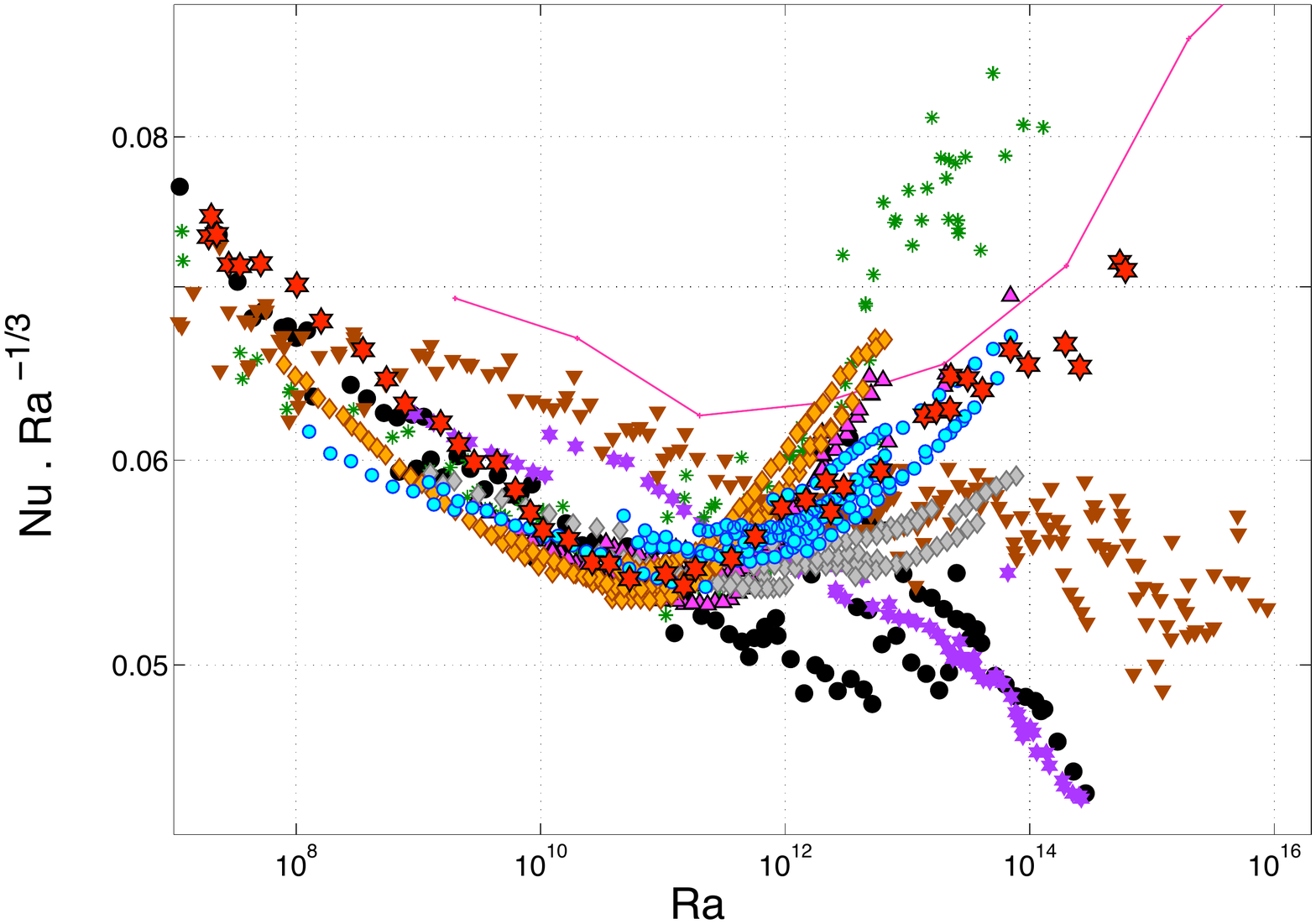}
\caption{Compensated heat transfer $Nu\cdot Ra^{-1/3}$ versus $Ra$ for very high $Ra$
experiments in cylindrical cell of aspect ratio $0.5 \le \Gamma \le 1.14$ and for $0.6<Pr<7$.
Datasets from Grenoble (green asterisks {[}\textit{Chavanne cell, $\Gamma=0.5$}{]}, magenta pointing-up triangles {[}\textit{Vintage cell, $\Gamma=0.5$}{]},
  blue disks {[}\textit{Flange
cell, $\Gamma=0.5$}{]}, grey diamonds {[}\textit{Paper cell, $\Gamma=0.5$}{]} , orange diamonds {[}\textit{Short-cell, $\Gamma=1.14$}{]},  Trieste [red stars, $\Gamma=1$],
Oregon [brown pointing-down triangles,  $\Gamma=0.5$], Göttingen [purple stars,  $\Gamma=0.5$] and Chicago after correction (see text) [black disks,  $\Gamma=0.5$]. The line corresponds to Delft T-RANS numerical simulations in an aspect ratio 8:8:1 cell.}
\label{Flo:Fig-NuRa-compil}
\end{figure}

Table \ref{table:manipHautsRa} summarizes the main specifications of Rayleigh-Bénard experiments reaching very high $Ra$. Bibliographic references are provided in the last column. For convenience, a name is attributed to each experiment performed in Grenoble ($2^{nd}$ column). Figure \ref{Flo:Fig-NuRa-compil} gathers measurements of the compensated heat transfer efficiency $Nu\cdot Ra^{-1/3}$ versus $Ra$. For completeness, a numerical simulation is included on the plot while -for clarity- a few experiments from Grenoble are omitted. The Chicago data have been re-calculated using improved He properties fits and corrected for a sidewall spurious effect (see \cite{RocheJLTP2004} for details on both corrections).
This figure clearly illustrates that below $Ra\sim 10^{11}$, all experiments are in reasonable agreement while different trends appear  above $Ra \sim 10^{11}$. Indeed, the compensated heat transfer $Nu\cdot Ra^{-1/3}$ decreases with $Ra$ in some experiments (Chicago, Oregon, Göttingen) while it increases in others (Grenoble, Trieste), leading up to nearly 100\% difference in heat transfer efficiency around $Ra=10^{14}$.  

The convection community doesn't agree on the description of the results at very high $Ra$. For example, a recent review on convection concludes ``\textit{Though the Grenoble experiments suggest such a transition near $Ra=10^{11}$ neither the Oregon-Trieste experiments nor numerical simulations do so. The reason for the discrepancy is presently unresolved} [...].'' \cite{AhlersGrossmannLohse_Review2009}, while we tend to consider that Trieste experiment (red star on figure \ref{Flo:Fig-NuRa-compil}) and Delft simulations (continuous line on figure \ref{Flo:Fig-NuRa-compil}) rather seem to fall into the group of transiting cells. Beside, we will argue in section \ref{sectionPr} that the cited experiments are not in disagreement.

The terminology used in the literature probably adds to the confusion. For example, the adjective ``ultimate" introduced by Chavanne \etal to name the new regime found in Grenoble, has been used in the literature to refer to four different items :  i) the regime observed in Grenoble (regardless of its interpretation), ii)  the concept of asymptotic regime of convection iii) Kraichnan's model, iv) a homogeneous turbulent flow forced by a thermal gradient \cite{lohse2003}. To avoid any confusion, we will avoid to refer to ``Ultimate regime'' or ``Ultimate state'', and we will call the ``Grenoble regime'' the regime which has been found and studied in Grenoble over the last 15 years.

The definitions of the Rayleigh, Prandtl and Nusselt numbers are :

\begin{center}
$Ra=\frac{\alpha \Delta h^3 g}{\kappa \nu}$, 
$Pr=\frac{\nu}{\kappa}$, 
$Nu=\frac{P}{P_{diff}}$
\end{center}

\noindent where $\alpha$, $\kappa$, $\nu$, $g$ and $h$ are respectively the isobaric thermal expansion coefficient, the molecular thermal diffusivity, the kinematic viscosity, the gravitational acceleration and the cell height. $P_{diff}$ is the power that would diffuse through the cell if the fluid was quiescent. The total power $P$ transported across the cell  and the temperature difference $\Delta$ driving the flow  are corrected to take the adiabatic gradient into account. We recall that this correction is exact.

\subsection{Motivation and Organisation of the paper}

The initial motivation of this paper are two questions hereafter. They are addressed varying independently the Rayleigh number (within $10^8<Ra<6.10^{14}$) and the Prandtl number (within $0.6<Pr<7$), in order to disentangle the influence of these two control parameters.

\begin{enumerate}
\item What is the nature of the Grenoble regime? In particular, is it the regime predicted by Kraichnan ?

\item What are the triggering conditions of the Grenoble regime?  As recalled later, Kraichnan's model doesn't describe the transition region, which leaves us with no precise prediction. %
\end{enumerate}

To address these questions, we performed a set of experiments which are described in section \ref{sectionManip}.
The following three sections explore the roles of the Prandtl number (section \ref{sectionPr}), the large scale circulation of the flow (section \ref{sectionLSC}) and the sidewall of the cell (section \ref{sectionSideWall}) in the triggering of the Grenoble regime. Section \ref{discu} summarizes the characteristics of this regime, and discusses them in connection with Kraichnan's model. Finally,  the Appendix presents a systematic study of non-Boussinesq effects, both experimental and theoretical.

\begin{table}
\caption{\label{table:manipHautsRa}Rayleigh-B\' enard experiments reaching very high $Ra$.}
\resizebox{!}{\textheight}{ %
\begin{sideways}
\begin{tabular}{|lc|c|c|cc|cc|cc|c|}
\hline 
\multicolumn{2}{|c|}{\textbf{Experiment}} & \textbf{Marked} & \textbf{Fluid} & \multicolumn{2}{c|}{\textbf{Cell}} & \multicolumn{2}{c|}{\textbf{Top / Bottom Plate}} & \multicolumn{2}{c|}{\textbf{Side wall }} & \textbf{Ref.}\tabularnewline
location & name &  \textbf{Transition$^0$} &  & height & aspect ratio $\Gamma$ & thickness  & material & thickness & material & \tabularnewline
 & \textit{comment} & $ Ra\in 10^{11-13}$ &  & $h$ {[}cm{]} & $\frac{\phi}{h}$ or $\frac{l}{h}:\frac{l}{h}:1$ & $h_{plate}$ {[}cm{]} &  & $e$ {[}mm{]} & & \tabularnewline
\hline
\multicolumn{1}{l}{} & \multicolumn{1}{c}{} & \multicolumn{1}{c}{} & \multicolumn{1}{c}{} &  & \multicolumn{1}{c}{} &  & \multicolumn{1}{c}{} &  & \multicolumn{1}{c}{} & \multicolumn{1}{c}{}\tabularnewline
\hline 
Grenoble & \textit{Chavanne} & \textbullet & He cryo & 20 &  0.5 & 2.5/2.5  & Cu/Cu & 0.5 & SS$^{(1)}$ & \cite{Chavanne1997,Chavanne2001,GauthierShotNoise:EPL2008}\tabularnewline
Grenoble & \textit{Corrugated} & \textbullet & He cryo & 20 & 0.5 & 2.5/2.5 $^{(2)}$ & Cu/Cu $^{(2)}$  & 0.5-0.6 $^{(2)}$  & SS$^{(1)}$ & \cite{RochePRE2001,RochePoF2005}\tabularnewline
Grenoble & \textit{Flange} & \textbullet & He cryo & 20 & 0.5 & 2.5/2.5 & Cu/Cu & 0.5 & SS$^{(1)}$ & present (see fig. \ref{Fig:cellules}-d)\tabularnewline
Grenoble & \textit{Paper} & \textbullet & He cryo & 20 & 0.5 & 2.5/2.5 & Cu/Cu & 0.16 + 0.5 & Paper+SS$^{(1)}$ & present\tabularnewline
Grenoble & \textit{Cigar} & \textbullet & He cryo & 43 & 0.23 & 2.5/2.5 & Cu/Cu & 0.5 & SS$^{(1)}$ & \cite{Salort:ETC12_2009}  (see fig. \ref{cigare}-b)\tabularnewline
Grenoble & \textit{Brass} & \textbullet & He cryo & 20 & 0.5 & 2.5/2.5 & Cu/Brass & 0.5   & SS$^{(1)}$ & \cite{RochePoF2005}\tabularnewline
Grenoble & \textit{Screen} & \textbullet & He cryo & 20 & 0.5 & 2.5/2.5 & Cu/Cu & 0.5 & SS$^{(1)}$ & present (see fig. \ref{Flo:Fig-LSC-1}-b)\tabularnewline
Grenoble & \textit{Vintage} $^{(3)}$ & \textbullet & He cryo & 20 & 0.5 & 2.5/2.5 & Cu/Cu & 0.5 & SS$^{(1)}$ & present (see fig. \ref{Flo:Fig-LSC-2}-b)\tabularnewline
Grenoble & \textit{CornerFlow} & \textbullet & He cryo & 20 & 0.5 & 2.5/2.5 & Cu/Cu & 0.5$^{(4)}$ & SS$^{(1)}$ & \cite{GauthierETC11_2007}\tabularnewline
Grenoble & \textit{ThickWall} & \textbullet & He cryo & 20 & 0.5 & 2.5/2.5 & Cu/Cu & $ $2.2 & SS$^{(1)}$ & \cite{GauthierETC11_2007}\tabularnewline
Grenoble & \textit{Short} & \textbullet & He cryo & 8.8 & 1.14 & 10/2.5 & Cu/Cu & 0.5 & SS$^{(1)}$ & present  (see fig. \ref{Fig:cellules}-a/b)\tabularnewline
\hline
\hline 
Chicago &  & - & He cryo & 40 & 0.5 & 6/2 & Cu/Cu & 1.5 & SS$^{(1)}$ & \cite{WuTHESE}\tabularnewline
\hline
\hline 
Rehovot & \textit{Non-Boussinesq}$^{(5)}$ & - & SF6 & 10.5 & 0.73:0.73:1 & 1.9/? & Sapphire/Ni & \multicolumn{2}{c|}{plexi / fluid / ?} & \cite{Ashkenazi1999}\tabularnewline
\hline
\hline 
Oregon &  & - & He cryo & 100 & 0.5 & 3.8/3.8 & Cu/Cu & 2.67 & SS$^{(1)}$ & \cite{Niemela2000}\tabularnewline
Trieste &  & \textbullet & He cryo & 50 & 1 & 3.8/3.8 & Cu/Cu & 0.17$ $$^{(6)}$+2.67 & Mylar+SS$^{(1)}$ & \cite{Niemela2003}\tabularnewline
Trieste & \textit{Non-Boussinesq}$^{(5)}$ & ? & He cryo & 12.5 & 4 & 3.8/3.8 & Cu/Cu & 2.67 & SS$^{(1)}$ & \cite{Niemela:2006}\tabularnewline
\hline
\hline 
Göttingen 2009 &  & - $^{(7)}$ & SF6 & 224 &  0.5 & 4/3.5+0.5+2.5 & Cu/Cu-i-Cu$^{(8)}$ & \multicolumn{2}{c|}{plexi / fluid / shield} & \cite{Ahlers:NJP2009}\tabularnewline
\hline
\hline 
Delft & \textit{TRANS simulation} & \textbullet & -$^{(9)}$ & - & 8:8:1 & - & - & - & - & \cite{Kenjeres2002}\tabularnewline
\hline
\multicolumn{1}{l}{} & \multicolumn{1}{c}{} & \multicolumn{1}{c}{} & \multicolumn{1}{c}{} &  & \multicolumn{1}{c}{} &  & \multicolumn{1}{c}{} &  & \multicolumn{1}{c}{} & \multicolumn{1}{c}{}\tabularnewline
\multicolumn{11}{l}{(0) The bullet indicates that a steep enhancement of heat transfer in found above a $Ra$ threshold within $10^{11-13}$ (with $Nu(Ra)$ scaling exponent significantly large than $1/3$).}\tabularnewline
\multicolumn{11}{l}{(1) SS: stainless steel.}\tabularnewline
\multicolumn{11}{l}{(2) The Cu plates had 0.11 mm deep grooves spaced by 0.44 mm. A corrugated bottom brass plate with 0.145 mm deep}\tabularnewline
\multicolumn{11}{l}{ grooves spaced by 0.45 mm has also been operated (see \cite{RochePoF2005}). The sidewall was 0.6 mm thick with 0.1 mm deep grooves in it.}\tabularnewline
\multicolumn{11}{l}{(3) The main difference between this cell and \textit{Chavanne-cell} is a 1.3 or 3.6 degrees tilt with respect to the vertical direction.}\tabularnewline
\multicolumn{11}{l}{(4) An adjustable heating (cooling) ring is varnished on the external side of the sidewall, right above (below) the bottom (top) plate. }\tabularnewline
\multicolumn{11}{l}{(5) The non-Boussinesqness of the data was acknowledged by the authors
of these measurements.}\tabularnewline
\multicolumn{11}{l}{(6) The Mylar sheet was epoxied. Two additional Mylar strips of height cover the ``flange region 
2.5 cm (thickness 0.167 mm) }\tabularnewline
\multicolumn{11}{l}{ just above the bottom plate and also below the top plate''.}\tabularnewline
\multicolumn{11}{l}{(7) Recent unpublished data obtained with this cell evidence improved heat transfer near $Ra\simeq 4.10^{13}$.}\tabularnewline
\multicolumn{11}{l}{(8) The bottom plate consists in a sandwich Cu-epoxy-Plexiglass-epoxy-Cu. An aluminium composite plate has also been used.}\tabularnewline
\multicolumn{11}{l}{(9) Boussinesq equations are simulated for $Pr=0.71$}\tabularnewline
\multicolumn{1}{l}{} & \multicolumn{1}{c}{} & \multicolumn{1}{c}{} & \multicolumn{1}{c}{} &  & \multicolumn{1}{c}{} &  & \multicolumn{1}{c}{} &  & \multicolumn{1}{c}{} & \multicolumn{1}{c}{}\tabularnewline
\end{tabular}
\end{sideways}}
\end{table}

\section{\label{sectionManip}New very high resolution cryogenic He experiments }

\begin{figure}
\centering
\subfloat[]{\includegraphics[height=0.25\textheight]{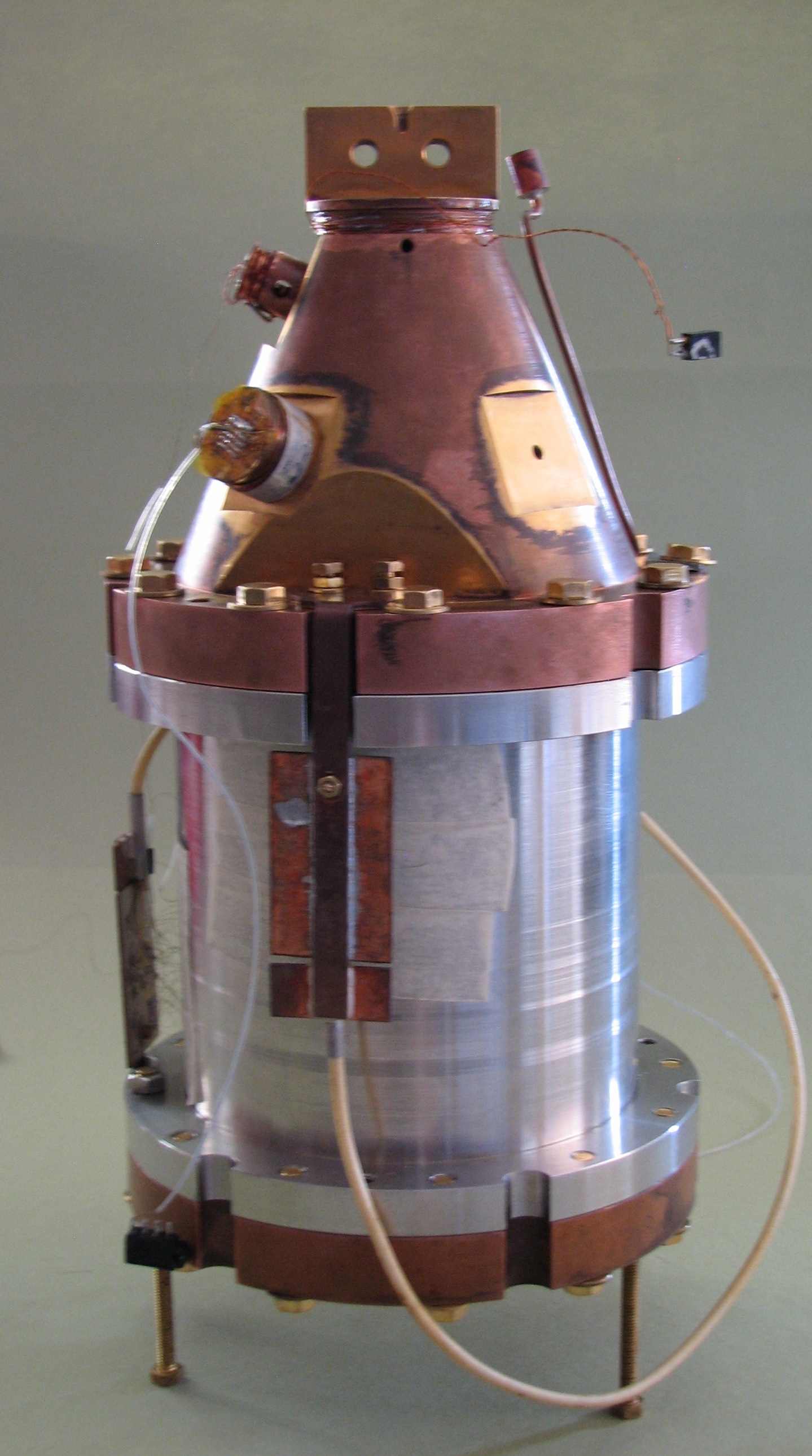}}
\subfloat[]{\includegraphics[height=0.25\textheight]{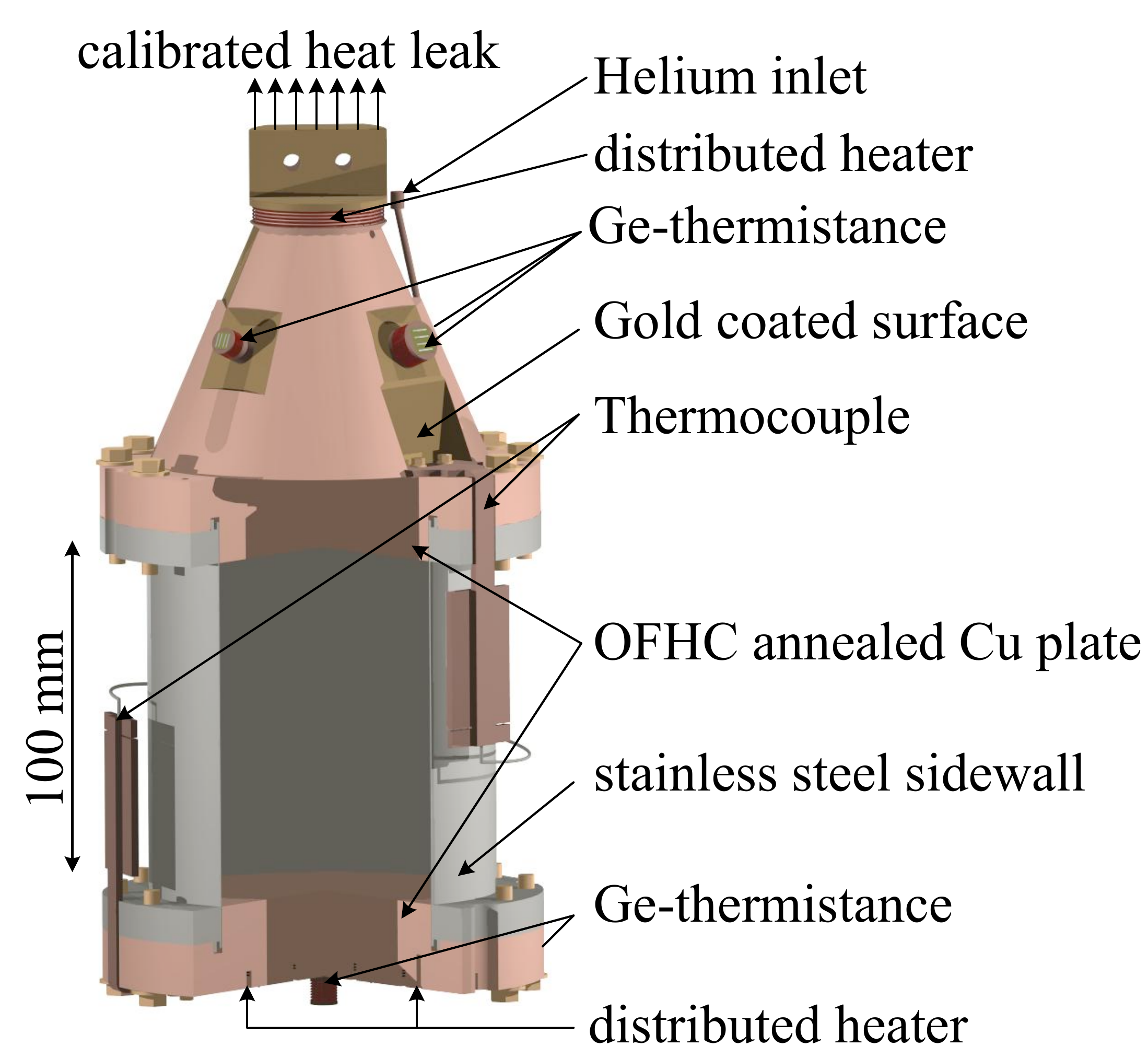}}
\subfloat[]{\includegraphics[height=0.25\textheight]{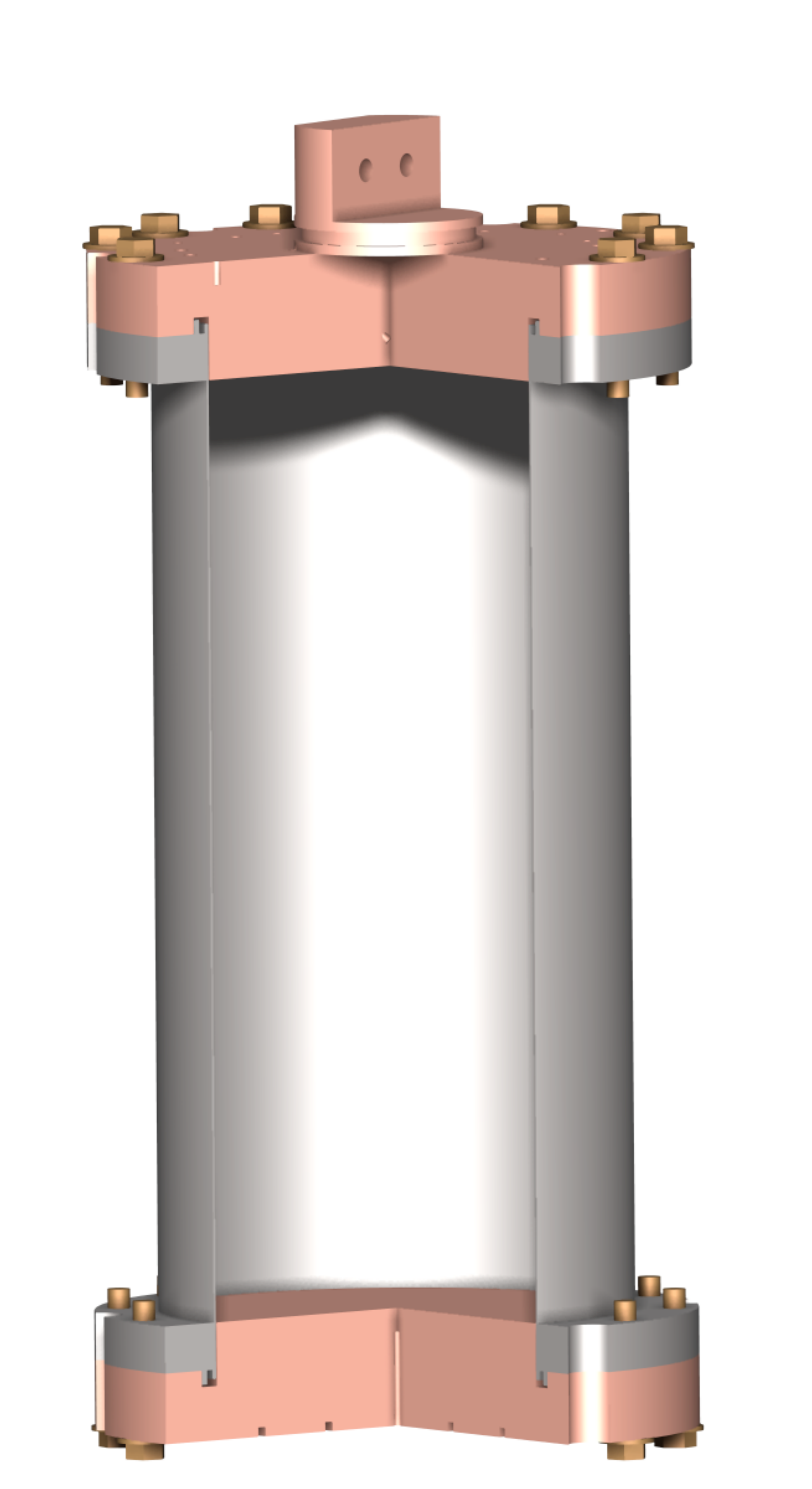}}
\subfloat[]{\includegraphics[height=0.25\textheight]{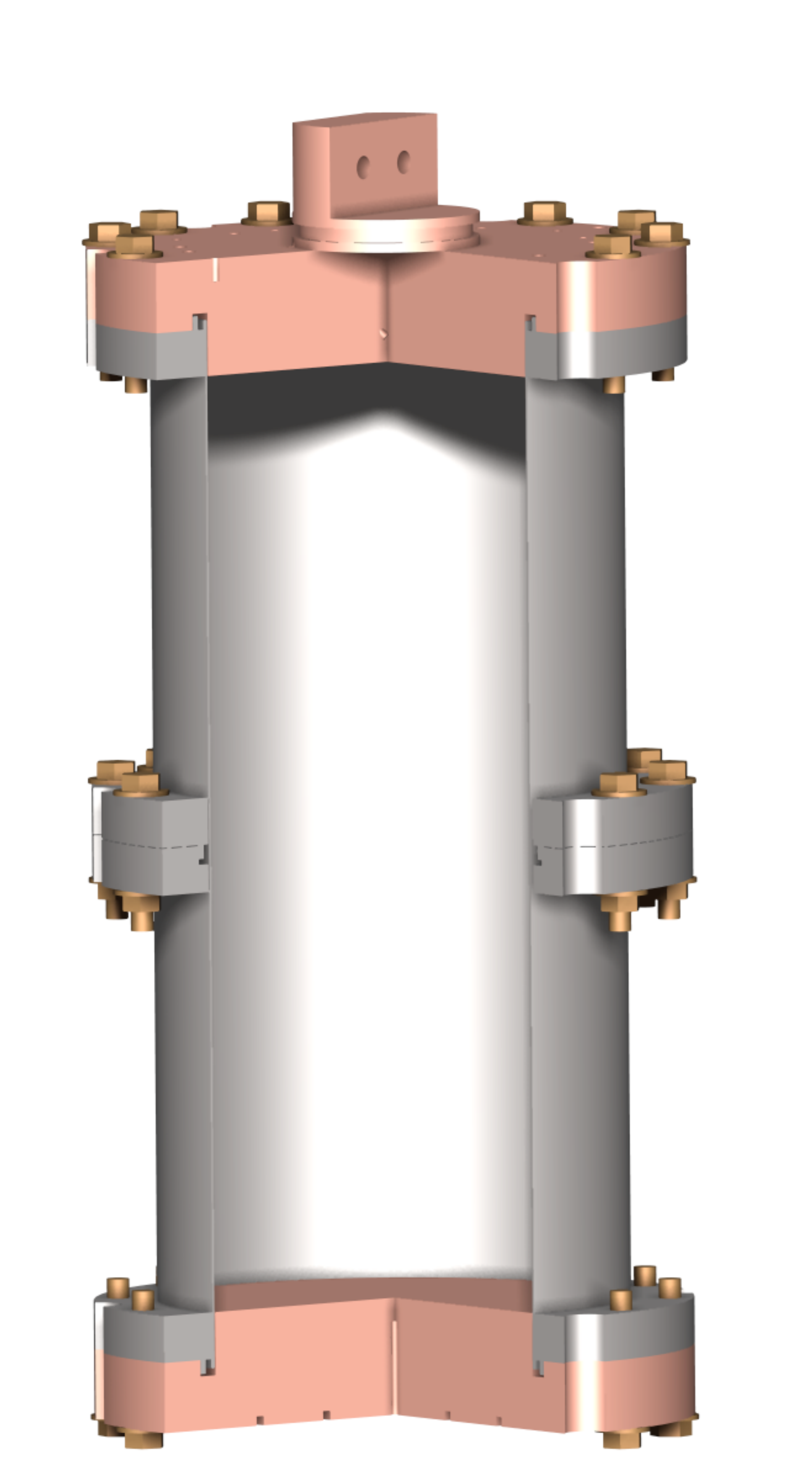}}
\caption{Cells before being hanged in a cryogenic-vacuum chamber : (a) and (b) Aspect ratio $\Gamma=1.14$ \textit{Short-cell}. (c) Aspect ratio $\Gamma=0.50$ typical cell. (d) \textit{Flangle-cell}.}
\label{Fig:cellules}
\end{figure}

\begin{figure}
\center
\includegraphics[width=.8\textwidth]{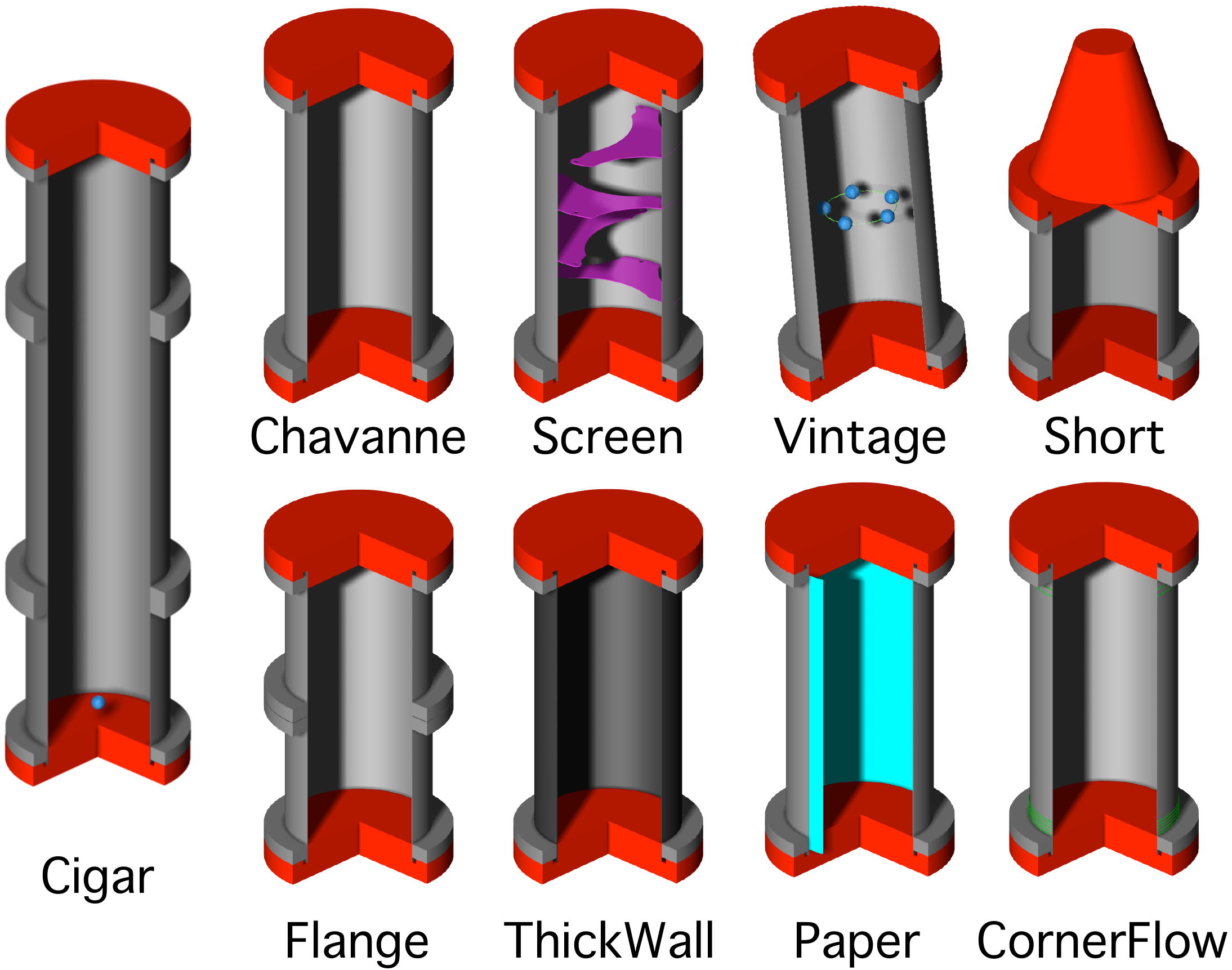}
\caption{Schematics of the Grenoble cells discussed in this work.}
\label{Flo:Fig-allCells}
\end{figure}

The seven cryogenic convection cells of present study are named the \textit{Flange}, \textit{Paper}, \textit{Cigar}, \textit{Screen}, \textit{Vintage}, \textit{ThickWall} and \textit{Short} cells.
All of them are cylindrical with diameter $\Phi=10\ cm$ and heights $h=8.8\,cm$ (\textit{Short-cell}), $43\,cm$ (\textit{Cigar-cell}) and $20\,cm$ (all the others), corresponding to aspect ratios $\Gamma \simeq 1.14,\,0.23$ and $0.50$ (see figures \ref{Fig:cellules} and \ref{Flo:Fig-allCells}).

The various top and bottom plates are $2.5$-cm-thick except for the conical top plate appearing on figure \ref{Fig:cellules}-a/b which is 10-cm-thick (\textit{Short-cell}). The conductivities of two Cu plates have been measured in-situ, as described in \cite{GauthierETC11_2007} and we found  $880$ and $1090\ Wm^{-1}K^{-1}$ at $4.2\ K$ for standard and OFHC Cu respectively. The other plates, most made of annealed OFHC Cu, are expected to have thermal conductivities of the same order. The heat capacity of the bottom plates (sidewall flange and screws included) has been measured to be $\sim 1\,J/K$. The measured roughness of all these Cu plates is typically $ra\simeq  0.15\, \mu m$ to $1\,\mu m$, depending on the cell, where $ra$ is the arithmetic average of the absolute vertical deviation (often noted $R_a$). The flatness of the surfaces in contact with the fluid was typically within $\pm 4\,\mu m$ for all cells, except for one which has a $15\,\mu m$ deep bump on a side. For the record, the non-corrugated brass plate used in a previous experiment \cite{RochePoF2005} had a roughness of $ra\simeq 1\,\mu m$ and a flatness within $\sim \pm 10\,\mu m$.

Several seamless stainless steel sidewalls have been used, with thicknesses $2.2\,mm$ (\textit{ThickWall cell}) and $500-550\ \mu m$ (all the other cells). The  thermal conductance of each sidewall was measured in-situ. The conductance of the \textit{Paper}, \textit{Screen} and \textit{Vintage} cells was $327\,\mu WK^{-1}$ at $4.7\,K$. The \textit{Flange}, \textit{Short}, \textit{ThickWall} and \textit{Cigar} cells conductances were respectively 1.15, 2.3, 4.8 and 0.5 times larger. The sidewall of the \textit{Flange cell} was assembled using two sidewalls of aspect ratio close to unity  (see figure \ref{Fig:cellules}-d); the motivation was to mimic the design of the Oregon cell. The parasitic contribution of sidewalls on $Nu$  was corrected using the analytical model from \cite{RocheEPJB2001},  confirmed in \cite{Verzicco2002}. This correction is very small at the very high $Ra$  of interest. For instance, around $Ra=10^{12}$,
the absolute value of the local scaling exponent of the heat transfer law $Nu(Ra)$ would be typically 0.01 larger without correction. The connection between the plate and the sidewall is detailed in \cite{GauthierETC11_2007} : it is such that the bottom (top) flange of the sidewall lies below (above) the bottom (top) plate-fluid interface.
 
The cells are hanging vertically in a cryogenic-grade vacuum, except for the \textit{Vintage cell} which was tilted by $1.3^{\circ}$ and $3.6^{\circ}$. The top plate is cooled by a helium bath at $4.2\ K$ through a calibrated thermal resistance (typically $2\ K/W$ at $6\ K$). The temperature is regulated by a PID controller. %
A constant and distributed Joule heating $P$  is delivered %
on the bottom plate. The heat leak from the bottom plate to the surrounding has been measured in-situ in a few experiments ($\simeq 200\ nW$ at $4.7\ K$) and it is three to four decades smaller than the lowest heating applied on the bottom plate to generate convection. This leak is mainly due to the radiative transfer to the environment at $4.2\,K$. This excellent thermal control is one of the advantages of our cryogenic environment over room temperature convection experiments, along with the excellent thermal properties of the Cu which provide isothermal plates up to the highest heat flux \cite{Verzicco_plate}.

The temperature difference $\Delta$ between the plates is measured with an accuracy down to $0.1\ mK$ thanks to specifically designed thermocouples. For comparison, the smallest $\Delta$ in our experiments are around $10\,mK$. The temperature of each plate is measured with various \textit{Ge} thermistances. Their calibration is checked in-situ against the critical temperature $T_c$ of the fluid with a resolution of $0.2\,mK$. To avoid a common misunderstanding, we stress that all the $Nu(Ra)$ measurements are done far away from the critical point, as argued in \ref{nob}. The critical point is simply used here as a thermodynamical reference to cross-check temperature calibration.

Cells are filled with various $^{4}He$ densities , ranging from dilute gas to liquid, and then closed with a cryogenic needle valve located close to the cell. The amount of \textit{He} introduced in the cell is measured in a calibrated tank at room temperature, and it is occasionally cross-checked at low temperature by measuring the condensation temperature. A thermosiphon hanging in vacuum between the valve and the cell prevents convective transfer in the filling line. \textit{He}  properties are calculated as described in \cite{RocheJLTP2004}. Most of the measurements are performed for temperature and densities where the accuracy on \textit{He} properties is the best, for example at $T=6\,K$ or around $\rho=70\,kg/m^3$ \cite{RocheJLTP2004}.

\begin{figure}
\center
\includegraphics[width=.8\textwidth]{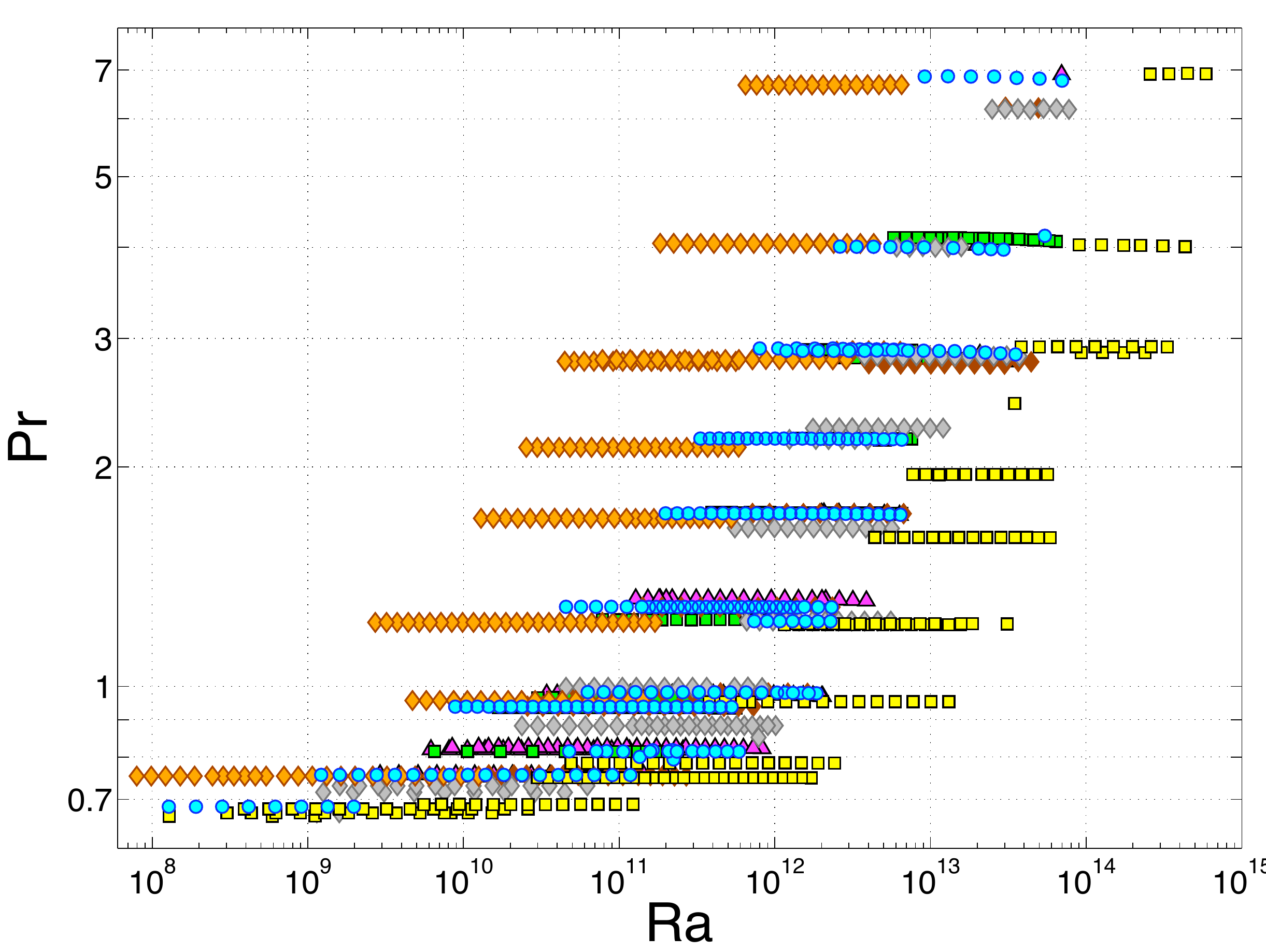}
\caption{Parameter space for the datasets \textit{Short} (orange diamonds), \textit{Vintage} (magenta pointing-up triangles), \textit{Flange} (blue disks), \textit{Paper} (grey diamonds), \textit{ThickWall} (green squares), \textit{Screen} (brown diamonds) \& \textit{Cigar} (yellow squares) cells.}
\label{Flo:Fig-RaPr-grenoble}
\end{figure}

\begin{figure}
\center
\includegraphics[width=.8\textwidth]{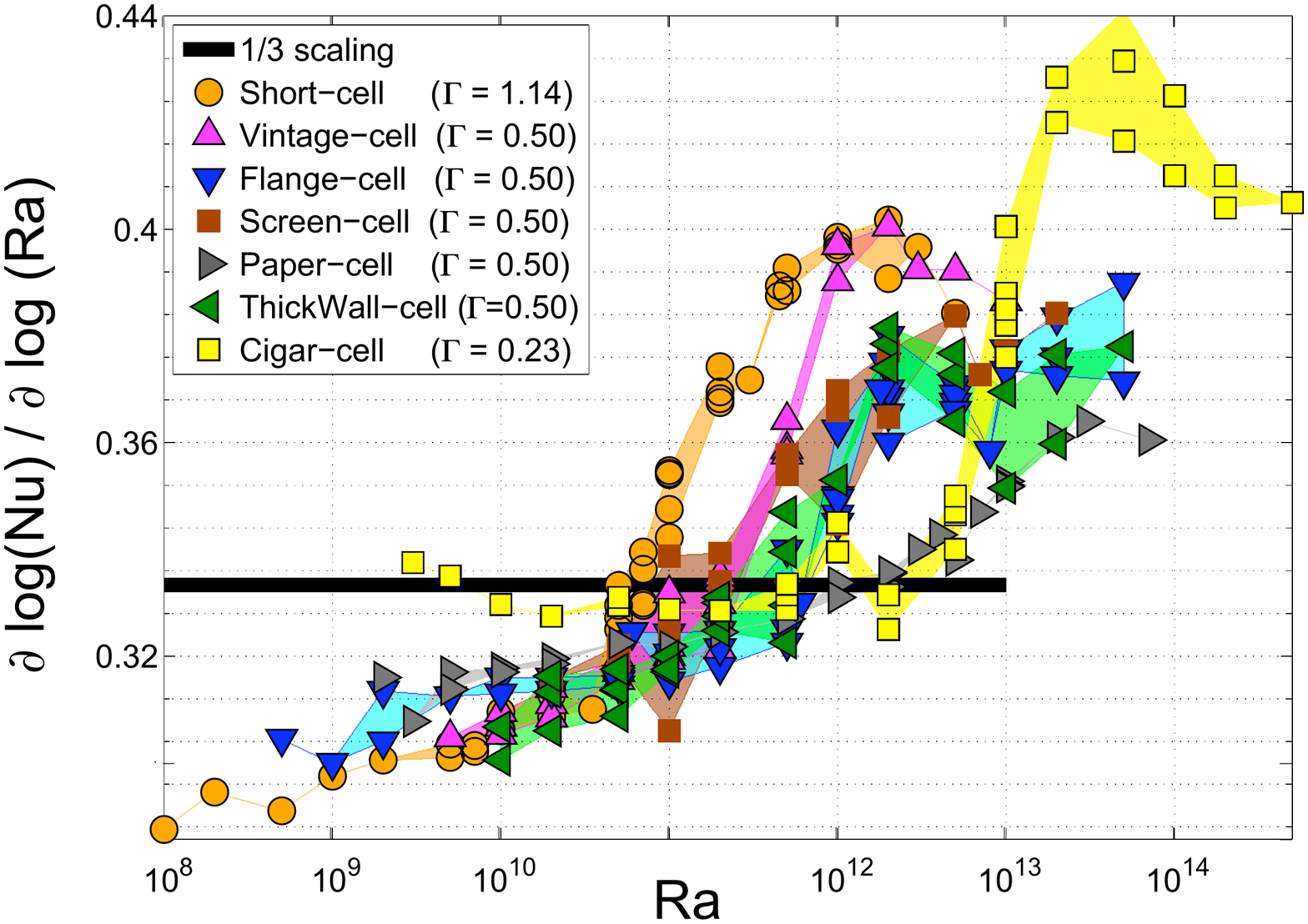}
\caption{Local exponent of the $Nu(Ra)$ law in various Rayleigh-Bénard cells for $0.6<Pr<7$.}
\label{Flo:Fig-exposants}
\end{figure}

Figure \ref{Flo:Fig-RaPr-grenoble} represent the \textit{Vintage}, \textit{Flange}, \textit{Paper}, \textit{Short}, \textit{ThickWall}, \textit{Screen} and \textit{Cigar} cells datasets in the $Ra-Pr$ parameter space. Contrary to a common misconception of cryogenic convection experiments, $Ra$ can be varied at a given $Pr$, as illustrated in this figure and already in previous works (e.g. \cite{WuTHESE,RocheEPL2002}). Each subset of constant $Pr$ data is obtained while working at a fixed mean temperature $T$ and mean volumetric mass $\rho$. For each of these subsets, the local scaling exponent of the $Nu(Ra)$ law -that is $\partial \log Nu / \partial \log Ra $-, can be determined with high accuracy because the uncertainty on the fluid properties prefactors appearing in $Nu$ and $Ra$ vanishes. The local exponents determined for each of these experiments are plotted on figure \ref{Flo:Fig-exposants}. On such a plot, the transition to the Grenoble regime can be easily spotted by the increase of the exponent above the 1/3 value. We underline the variability of the transition, both in term of transitional $Ra$ (nearly two decades) and in terms of strength with exponent from 0.36 up to 0.44 at $Ra\simeq 10^{14}$

\section{\label{sectionPr}Grenoble regime and Prandtl number}

This section compiles existing measurements and report new ones on the interplay between the Prandtl number $Pr$ and the Grenoble regime.

\subsection{Grenoble regime in the $Ra-Pr$ parameter space.}

\begin{figure}
\begin{minipage}[b][][t]{0.7\textwidth}
\includegraphics[width=.95\textwidth]{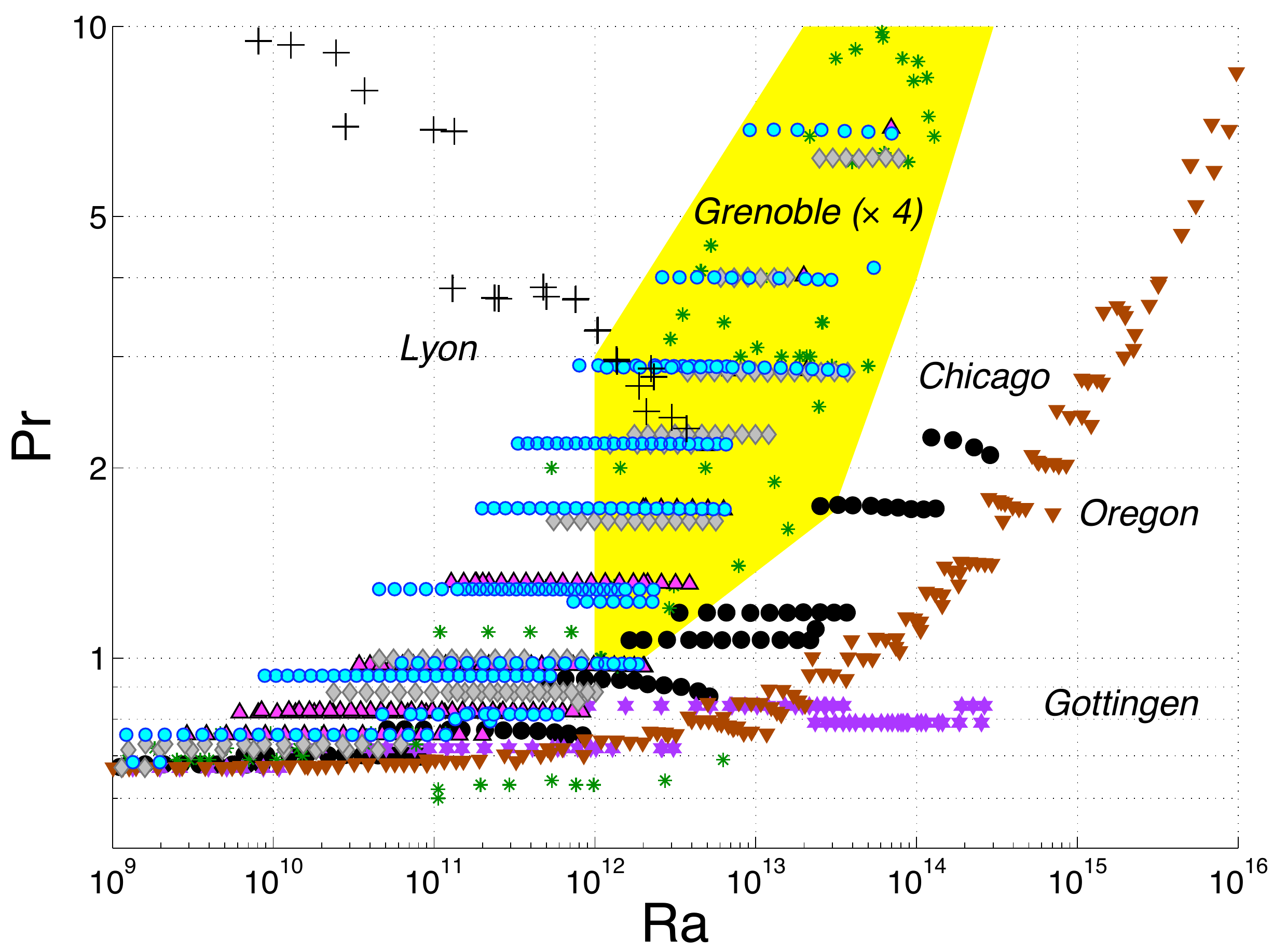} (a)
\end{minipage}
\begin{minipage}[b][][t]{0.3\textwidth}
\centering
 \includegraphics[height=0.15\textheight]{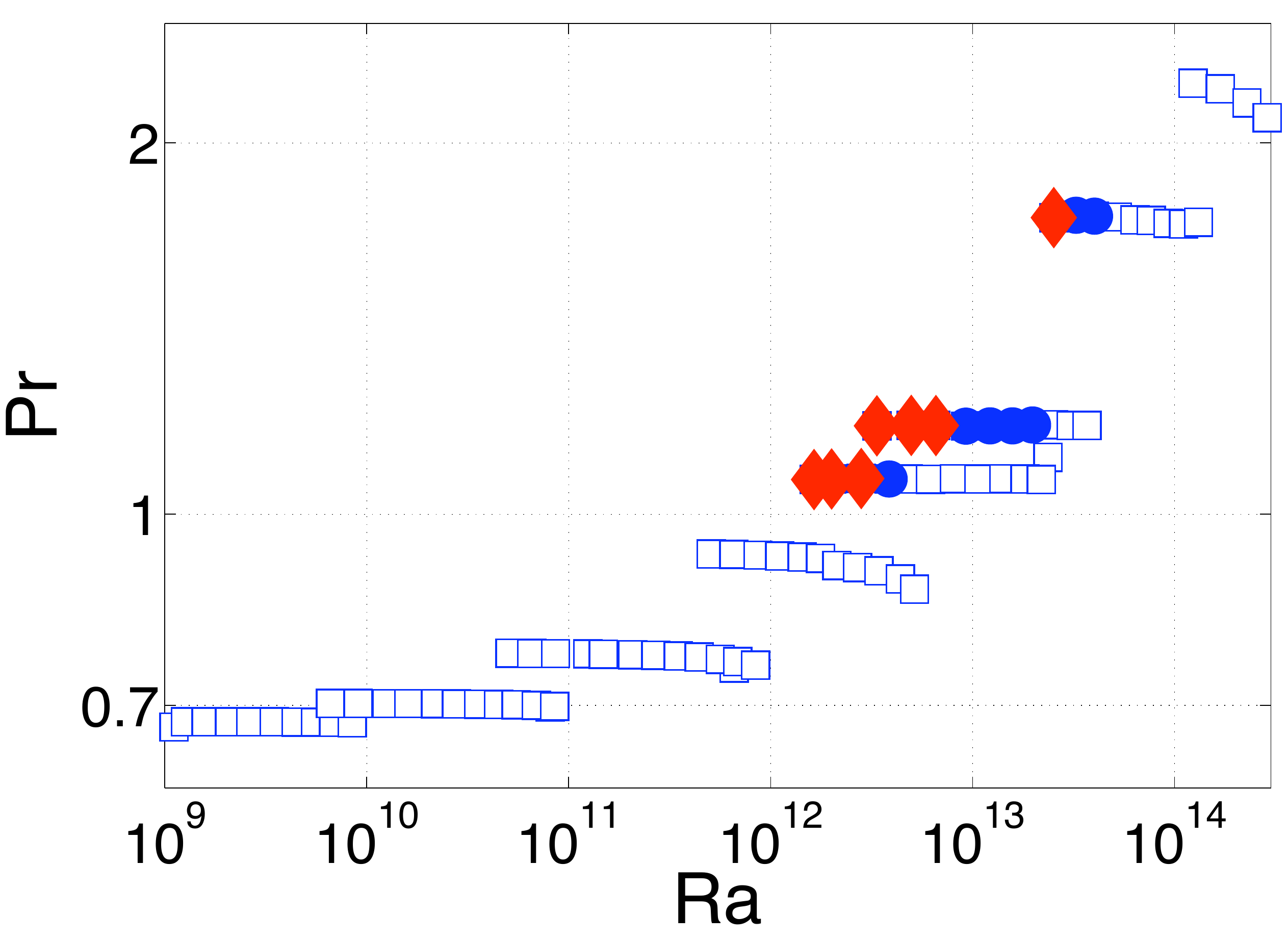} 
 \\(b)
\includegraphics[height=0.15\textheight]{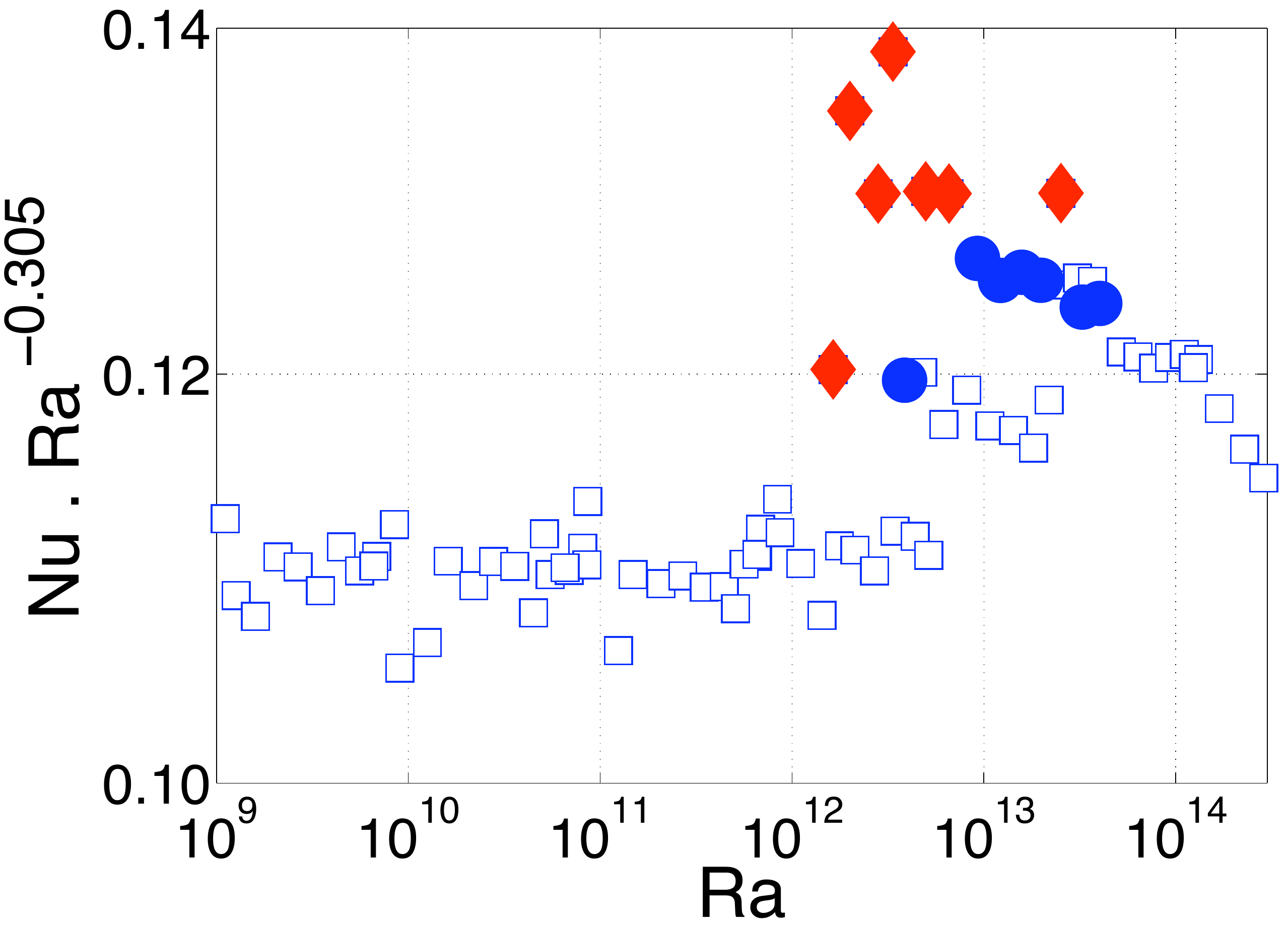} (c)
\end{minipage}
\caption{Left: Parameter space of very high $Ra$ experiments with aspect ratio $\Gamma=0.5$. Right : Chicago dataset. The points which are the closest to the yellow area are marked with full symbols (blue disks and red diamonds for the closest ones), while the others are plotted with open squares.}
\label{Flo:Fig-RaPr-Gamma05}
\end{figure}

Figure \ref{Flo:Fig-RaPr-Gamma05}-a  gathers various very high $Ra$ measurements in cylindrical aspect ratio $\Gamma=0.5$ Rayleigh-Bénard cells. The yellow area encircles all the measurements taken in the Grenoble regime. Reversely, nearly all the datasets reaching very high $Ra$ without experiencing a (clear) transition are falling outside this area. A few points from  Chicago are falling close to this area. Very interestingly, these points seems to experience a small heat transfer enhancement as shown on the figures \ref{Flo:Fig-RaPr-Gamma05} (right side subplots).  Another interesting point is Chavanne \etal points for $0.6<Pr<0.7$  and $10^{11}<Ra<10^{13}$: no heat transfer enhancement is distinguishable on these points, which is consistent with the location in the $Ra-Pr$ parameter space. As a first point, it is worth stressing that the cryogenic data from Grenoble, Chicago and Oregon are not in disagreement as long as differences in $Pr$ are regarded. As a second point, in addition to the very high $Ra$ condition, the transition is favoured above a $Pr(Ra)$ threshold. The only possible contradiction between these very high $Ra$ datasets may be found around $Pr\sim 2-3$, with the Lyon water experiment\cite{Chaumat:ETC9_2002}, which reported no transition. The corresponding dataset seems in contradiction over half a decade of $Ra$, right above the transition. We will show that the transition threshold can greatly vary from one cell to another depending on the sidewall boundary conditions (section \ref{sectionSideWall}). This may explain the apparent contradiction. Another possible explanation are accuracy issues at the highest $Ra$ in room temperature experiments due to significant correction of plate effects, which could mask the onset of a transition. Further investigations in this region of the parameter space, with smaller \textit{He} cells and larger water ones, would be useful.

 Unfortunately, by lack of available experiments, it is not yet possible to generalise this $Ra-Pr$ phase-space description to cells with other aspect ratios and shapes. %
 However, we will show in section \ref{sectionSideWall} that the transition $Ra$ has a strong aspect-ratio dependence in the range $0.23 \leq \Gamma \leq 1.14$, at least for $Pr$ of order one. 

\subsection{\label{PrSubSec}Prandlt number dependence of the Grenoble regime}

\begin{figure}
\centering
\subfloat[]{\includegraphics[height=0.2\textheight]{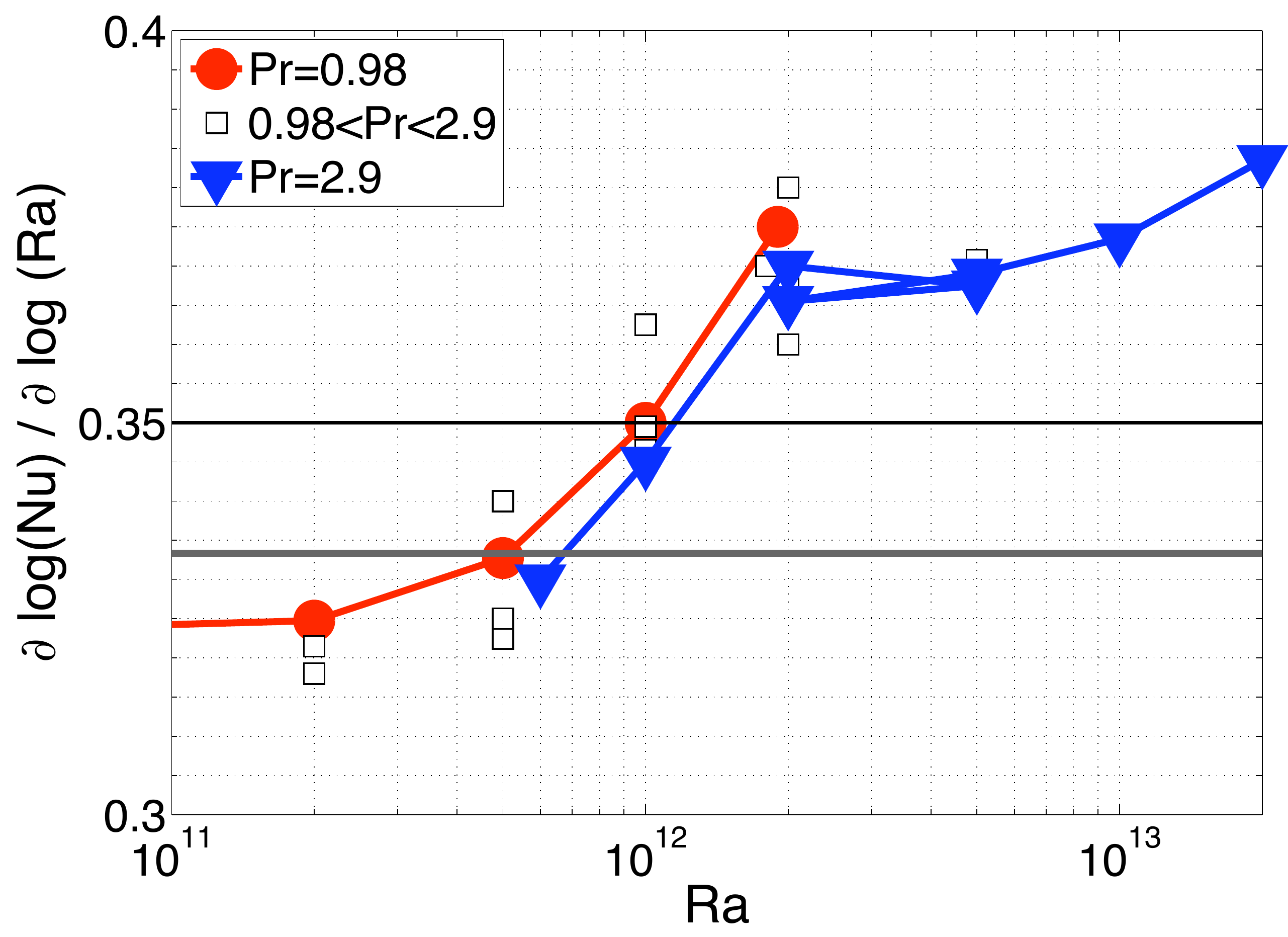}}
\qquad
\subfloat[]{\includegraphics[height=0.2\textheight]{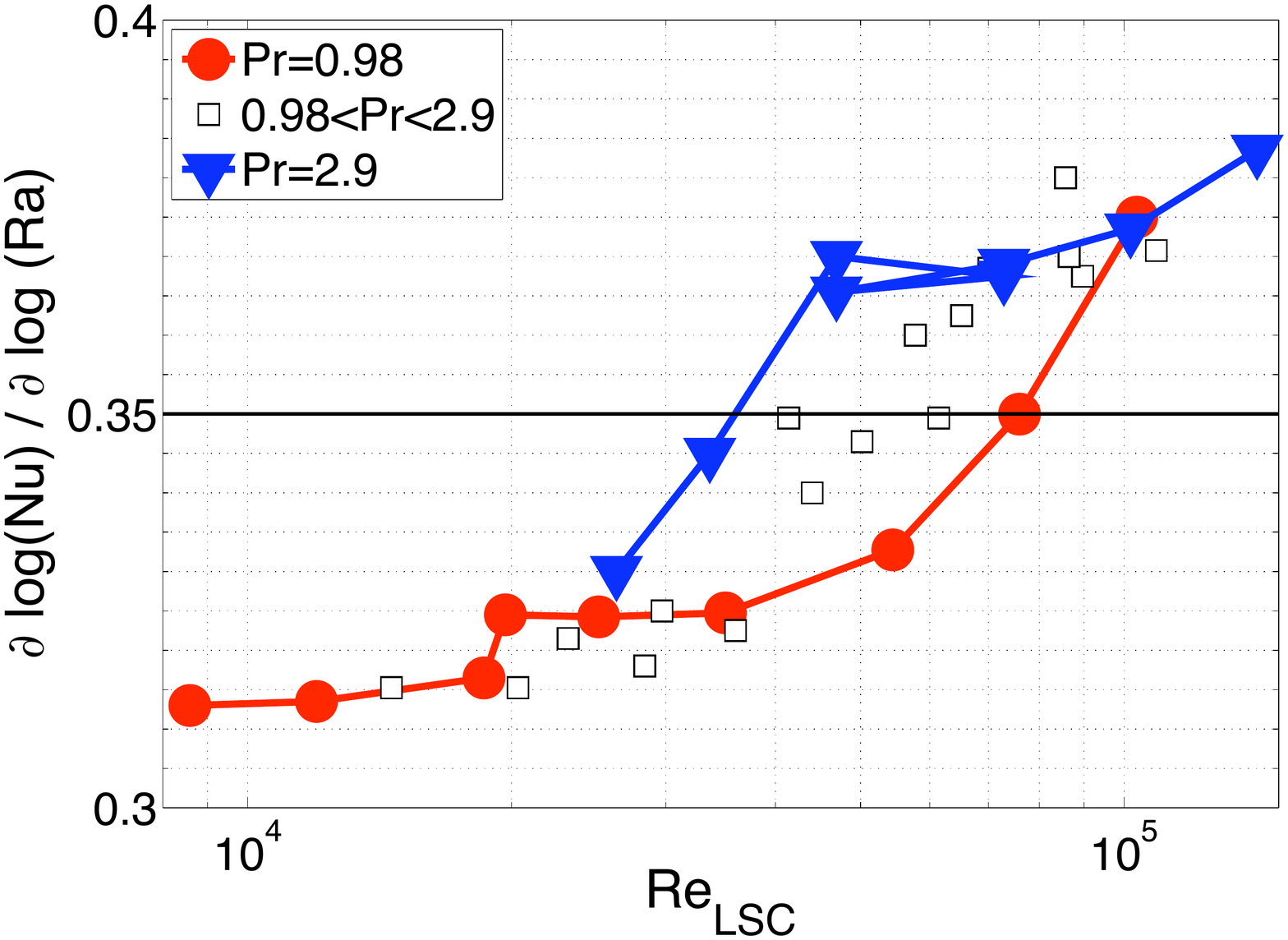}}
\caption{Local exponent of the $Nu(Ra)$ for various $Pr$ in the  \textit{Flange cell} ($\Gamma=0.50$). (a) versus $Ra$ - (b) versus the Reynolds number of the large scale circulation. 
}
\label{Flo:Fig-Pr}
\end{figure}

Does the $Ra$ of the onset of the transition depend on $Pr$ ? Figure \ref{Flo:Fig-Pr}-a gives the local scaling exponent of $Nu(Ra)$ around the transition within $0.98 \leq Pr \leq 2.9$. %
As can been seen, there is no $Pr$-dependence of the transition $Ra$, within accuracy, in this limited range of $Pr$.

\begin{figure}
\centering
\subfloat[]{\includegraphics[height=0.2\textheight]{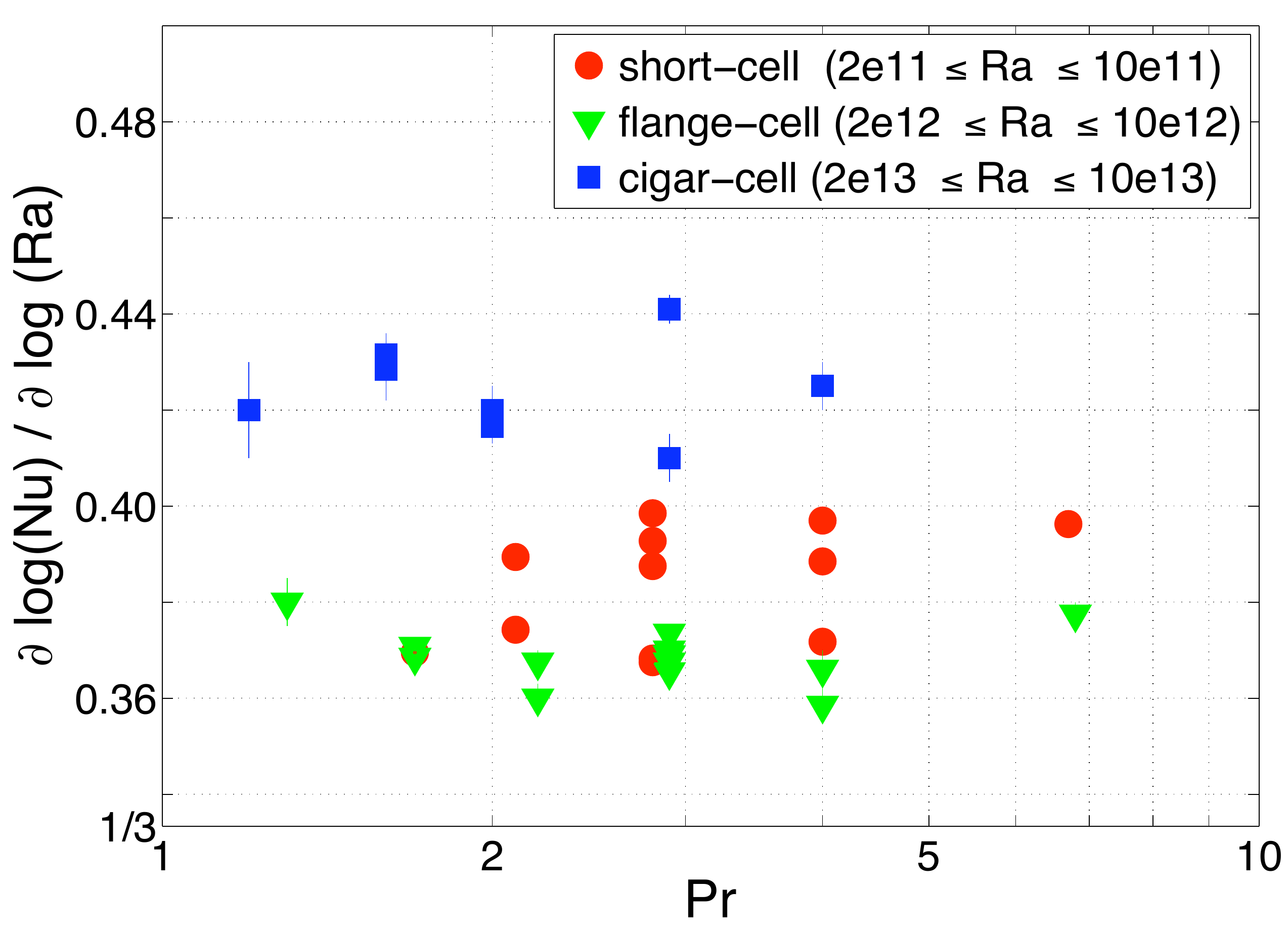}}
\qquad
\subfloat[]{\includegraphics[height=0.2\textheight]{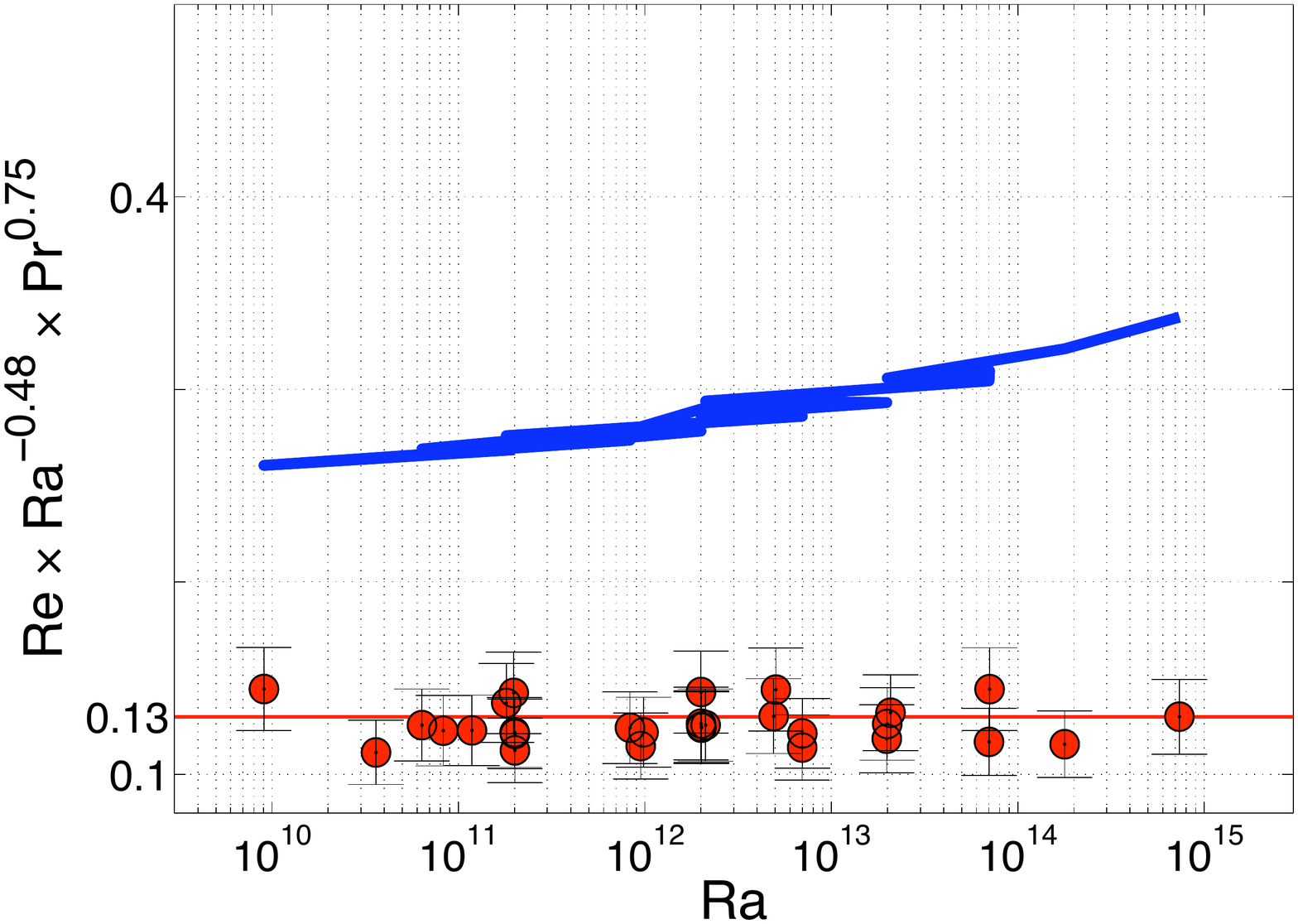}}
\caption{(a)  Local exponent of the $Nu(Ra)$ law versus $Pr$ for the highest $Ra$ in cell of aspect ratio $\Gamma=1.14$ , $0.50$ and $0.23$
(b) Compensated $Re$ associated with the Large Scale Circulation
for $\Gamma=0.50$. Red discs: based on $Re_{LSC}$. Blue segments: $Re$ based on  a local velocity fit at mid-height
(and at a distance $\Phi/4$ from the cell axis) measured previously in the same cell \cite{Chavanne2001}.
}
\label{Flo:Fig-LSC-3}
\end{figure}

As we will show later, the Reynolds numbers associated with the large scale circulation (LSC) in our $\Gamma=0.5$ cells roughly scale like $Re_{LSC} \simeq 0.13 Ra^{0.5} Pr^{-0.75}$ (see figure \ref{Flo:Fig-LSC-3}). If the transition was triggered when $Re_{LSC} $ reaches a critical value, a threefold increase of $Pr$ would shift the transitional $Ra$ by a factor $3^{0.75/0.48}\simeq 5.6$. This is not compatible with the result presented on figure \ref{Flo:Fig-Pr}-a which indicates a weaker -if any- dependence. Figure \ref{Flo:Fig-Pr}-b presents the same exponents versus $Re_{LSC}$ (estimated with the above fit) and shows that the transition does not occur for a unique value of $Re_{LSC}$.  A similar conclusion can be drawn for the aspect ratio $\Gamma=1.14$ cell over the same range of $Pr$. This suggests that the transition is not simply triggered by the LSC shear on the boundary layers. We will confirm this important result later. %

Figure \ref{Flo:Fig-LSC-3}-a shows the $Nu(Ra)$ local scaling exponent versus $Pr$ over a limited window of very high $Ra$, and for three cells with aspect ratio from 0.23 to 1.14. The $Ra$-window is chosen significantly above the transitional $Ra$ of each cell, with the purpose to be in the region where the new regime is well established. No systematic dependence of the local exponent versus $Pr$ is detectable within uncertainty within $1.2\leq  Pr \leq 6.8$. The best fit for $Nu(Pr)$ at constant and high $Ra$ gives $Nu\sim Pr^{0.045}$.

As a conclusion on this section, a first requirement for the occurrence of a transition is that the Rayleigh number is above a $Pr$-independent threshold.  In aspect ratio $\Gamma = 0.5$ cells, a second requirement has been evidenced :  $Pr$ should be above a $Ra$-dependent threshold. Both requirements are summarized by the phase space of figure \ref{Flo:Fig-RaPr-Gamma05}. Once the transition has occurred, the Prandtl number is found to have a small -if any- effect on heat transfer. We recall that our analysis focuses on the range $1 \lesssim Pr   \lesssim 7$ : it is therefore possible that this window of Prandtl numbers sits near the frontier between a low Prandtl number regime and a high Prandtl number regime.

\section{\label{sectionLSC}Grenoble regime and the Large Scale Circulation}

A large scale circulation (LSC) is often present in turbulent RB convection \cite{krishnamurti1981}.
This ``wind'' has a complex dynamics with multistability, reversals
and quiet periods. One of the oldest hypothesis to explain the puzzle
at very high $Ra$ was to invoke an interplay with the LSC (\cite{RocheTHESE},
p70). We report below a few tests of the interplay between the wind and the Grenoble regime.

\subsection{Is the wind changing when the transition occurs?}

The correlation between the LSC and Grenoble
regime can be assessed comparing the statistics of the LSC below and above the
transition. To probe the LSC, five thermometers \cite{Mitin2007} are suspended within
the flow, in the horizontal mid-plane of the \textit{Vintage-cell}. They are evenly distributed
on the circle which is equidistant from the cell axis and the sidewall (see figure \ref{Flo:Fig-LSC-2}-b).
Temperature time series are recorded simultaneously from the five thermometers.
The Reynolds number associated with the
LSC was determined using the auto-correlation technique. A LSC turn-over
time scale $T_{LSC}$ is defined imposing that the auto-correlation
of one of the thermometers (the most sensitive one was used) has its first minimum
at $T_{LSC}/2$. Physically, this characteristic time scale can be understood as the time of flight of the large scale temperature heterogeneities from one side of cell to the other side. %
The $Re$ associated with this LSC time is then defined as :

\begin{center}
$Re_{LSC}=\frac{2h^{2}}{\nu\cdot T_{LSC}}$ 
\par\end{center}

Figure \ref{Flo:Fig-LSC-3}-b shows $Re_{LSC}$ over five decades of $Ra$
and within $0.76\leq Pr\leq6.88$. We stress that $Ra$ and $Pr$ are varied independently, as can be seen on figure \ref{Flo:Fig-RaPr-grenoble}, which enables to fit independently the exponents of both $Ra$ and $Pr$. The compensation chosen on the y-axis illustrates the best power law fit:

\begin{center}
$Re_{LSC} \simeq \left( 0.13\pm0.03 \right) \times Ra^{0.48\pm0.02}\times Pr^{-0.75\pm0.03}$
\par\end{center}

This fit is in agreement with other ones that have been reported in the literature at lower $Ra$, in the hard turbulence regime
(e.g. \cite{Lam2002,Sun_PRE2005,brown2007}).  The important result here is that this fit also remains valid in the Grenoble regime.
In particular, no discontinuity is detectable at the transition. 

The Reynolds number $Re_{local}$ based on a previous local velocity measurement at mid-height in a similar cell is also plotted on figure \ref{Flo:Fig-LSC-3} \cite{Chavanne2001}.  The difference in magnitude and scaling between $Re_{LSC}$ and $Re_{local}$ is consistent with the differences in their definition (e.g. \cite{Sun_PRE2005}) \footnote{The local velocity measurement of was performed using two thermometers, one above the other, and by inferring the velocity from the cross-correlation of the two temperature time series \cite{Chavanne2001}. %
The slight difference in scaling ($Re_{LSC}\sim Ra^{0.48}$ versus $Re_{local} \sim Ra^{0.49}$) can be understood easily  the turn-over time scale $T_{LSC}$ is seen as the ratio of an effective LSC path length over its effective velocity. If we imagine that the effective path lengh of the LSC slightly evolves, we immediately find that $Re_{LSC}$ and $Re_{local}$ should have slightly different scalings \cite{Niemela_EPL2003}.} This previous determination of a LSC Reynolds number, as well as a third one done in Trieste (transiting) cell \cite{Niemela2001} also confirm that the strength of the LSC has no discontinuity when the transition occurs.

To complete the characterisation of the LSC, the statistics of its angular direction or ``polarisation" is measured.
At each time step of the temperature time series, the temperature distribution
along the rack of five probes is Fourier transformed versus angular position. The strength and polarisation of the LSC versus
time is defined from the amplitude and phase of the first Fourier
mode. Similar multi-probe techniques has been validated to analyse the LSC in previous studies (e.g. \cite{brown2006,XiXia:2008}). When the amplitude of the first mode is smaller than 1.3 times the rms amplitude of higher modes, we consider the LSC as undistinguishable from the background fluctuations. In this case, the LSC is considered as undefined.
Systematic measurements of the probability density functions (pdf)
of the LSC polarisation are done from $Ra=3.6\cdot10^{10}$ up to $Ra=2\cdot10^{12}$
(and $Pr\simeq1$). Examples of such  pdf are displayed on figure \ref{Flo:Fig-LSC-2}-a.
No significant discontinuity of these pdf is detectable at the transition,
in particular on the shapes and maximum value of the pdf. Similarly,
the fraction of time during which the LSC is considered as undefined
remains constant ($29\%\pm5\%$) over the whole range of $Ra$.

As a first conclusion about the wind, we showed that the occurrence of the transition is not associated with a discontinuity in strength, scaling nor in polarisation of the LSC.
One consequence is that the transition is not triggered by some instability in the dynamics of the LSC. Another consequence is that the transition does not alter significantly the LSC. 

\subsection{Is the transition altered by modifications of the large scale circulation ?}

The conclusion above does not rule out the possibility that the transition to the Grenoble regime is triggered by the LSC, in particular by its  strengthening with $Ra$. To explore how the transition depends on the LSC, we first altered it introducing some external constrains on the flow.

\begin{figure}
\centering
\subfloat[]{\includegraphics[height=0.22\textheight]{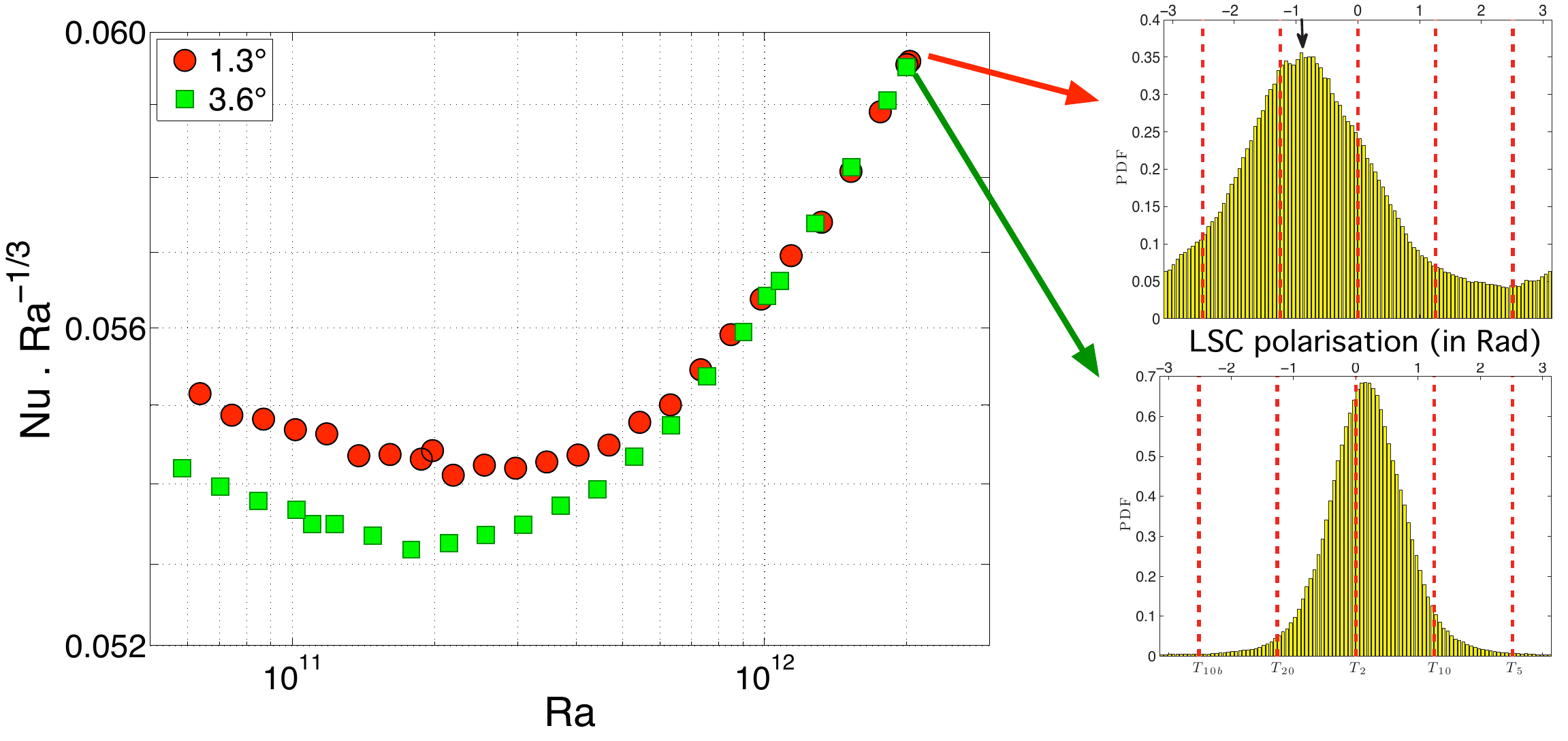}}
\qquad
\subfloat[]{\includegraphics[height=0.18\textheight]{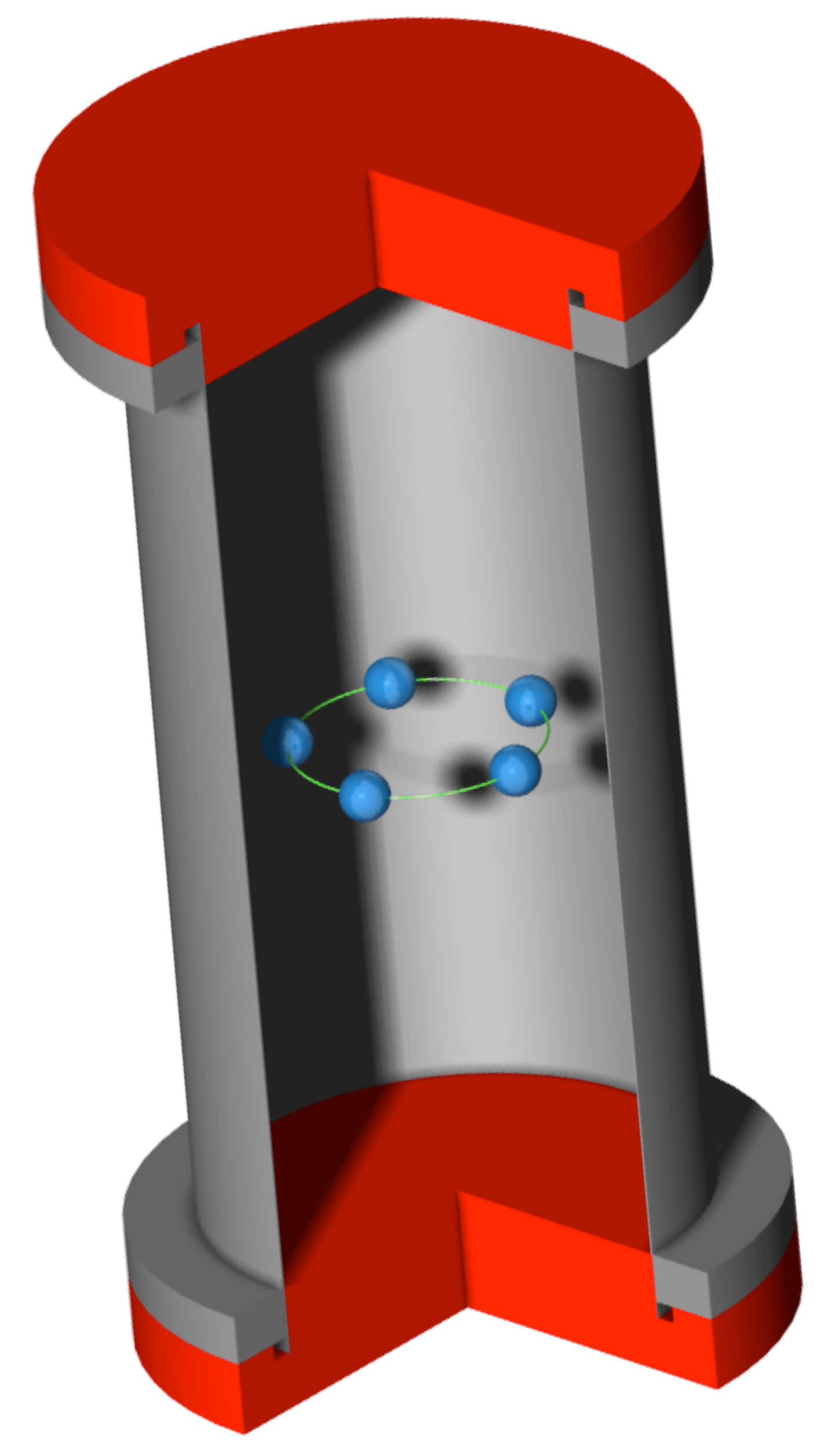}}
\caption{(a) Heat transfer (left figure) and probability density functions of the
LSC angular direction at $Ra=2\cdot10^{12}$ in the same
cell with a 1.3 degree tilt (discs and upper pdf) and with
a $3.60$ degrees tilt in a different direction (squares and lower pdf)
for $Pr=1.0$.
 (b) Schematic showing the probes positions in the \textit{Vintage} cell. The probe sizes and the tilt are exaggerated for visibility.}
\label{Flo:Fig-LSC-2}
\end{figure}

A first test consists in tilting the \textit{Vintage-cell}  from $1.3$ degree to $3.60\pm0.15$ degrees in a different direction.
The set of thermometers inside the flow (see previous subsection) confirms
that the LSC polarisation follows the new tilt direction and that its angular distribution becomes
sharper when the tilt angle is larger, as shown by the pdf of figure \ref{Flo:Fig-LSC-2}-a. Below the transition, increasing tilt reduces heat transfer
by 2\%, consistently with other experiments with the same aspect ratio
\cite{Chilla2004_TiltLongRelax,sun2005c}. On the contrary,
the heat transfer is unchanged in the Grenoble regime.
This robustness of the heat transfer to a tilt increase (and to a more pronounced polarisation of the LSC) is a new signature of the Grenoble regime. %

\begin{figure}
\subfloat[]{\includegraphics[height=0.22\textheight]{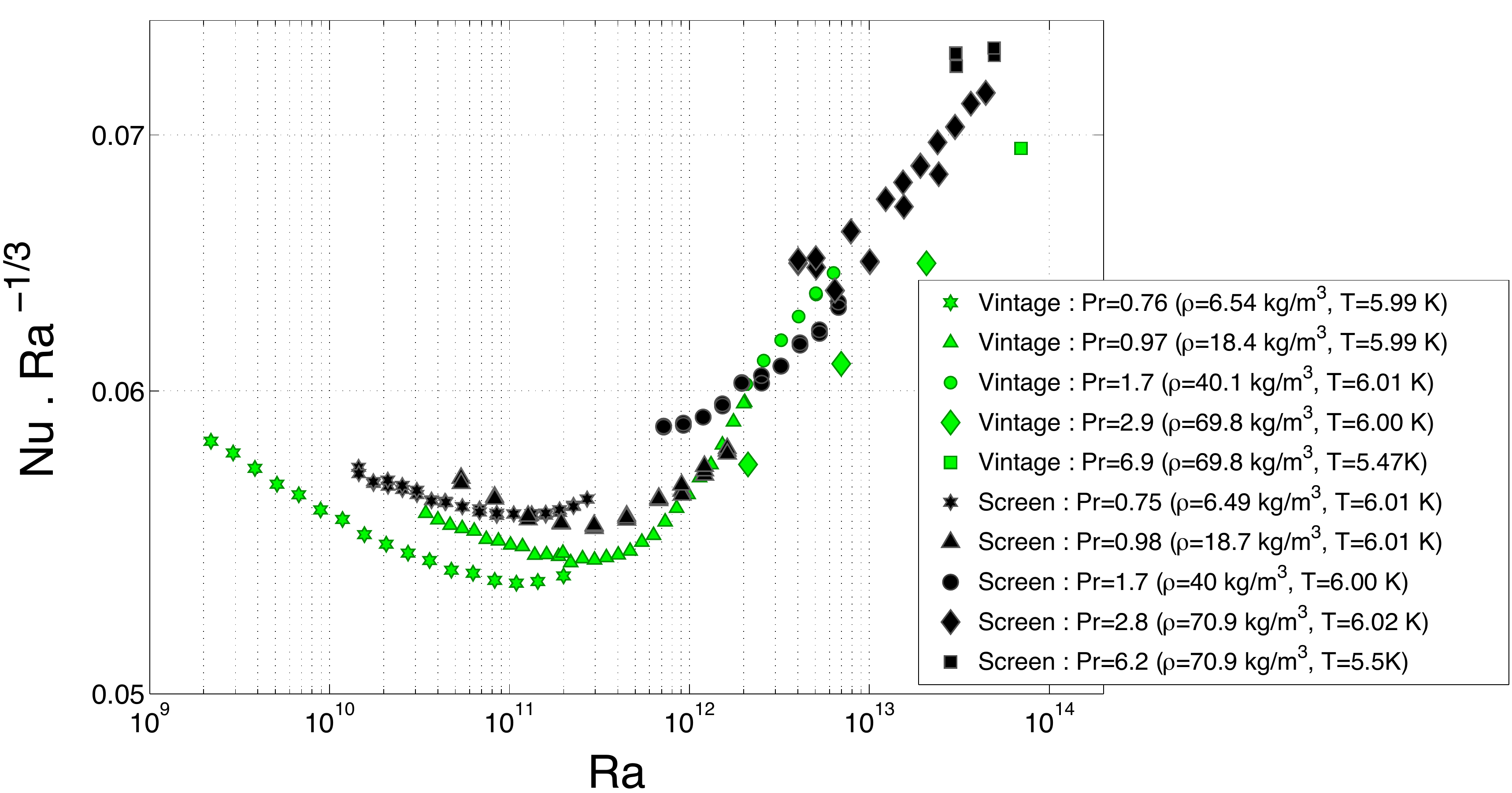}}
\subfloat[]{\includegraphics[height=0.18\textheight]{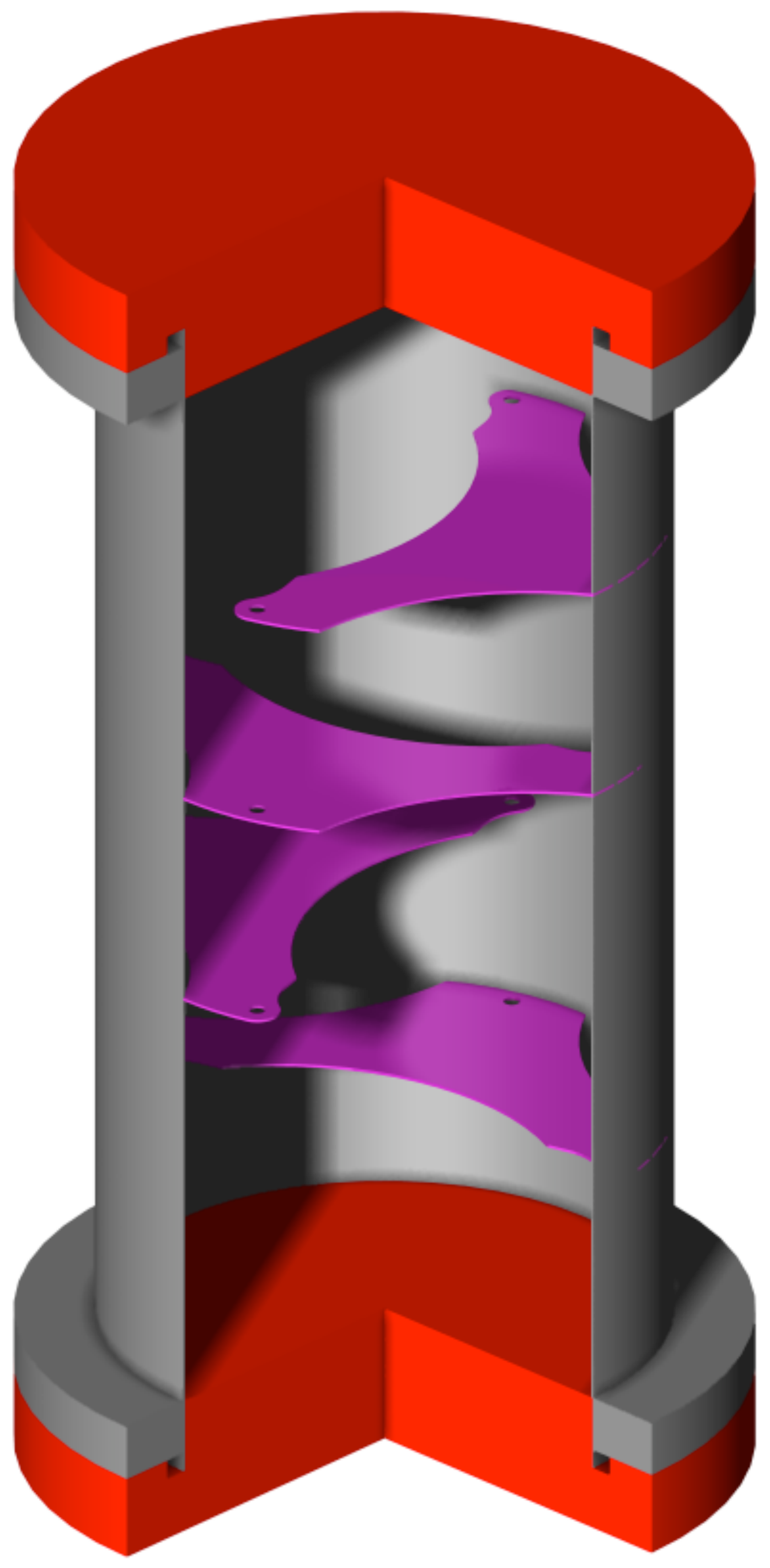}}
\qquad
\subfloat[]{\includegraphics[height=0.18\textheight]{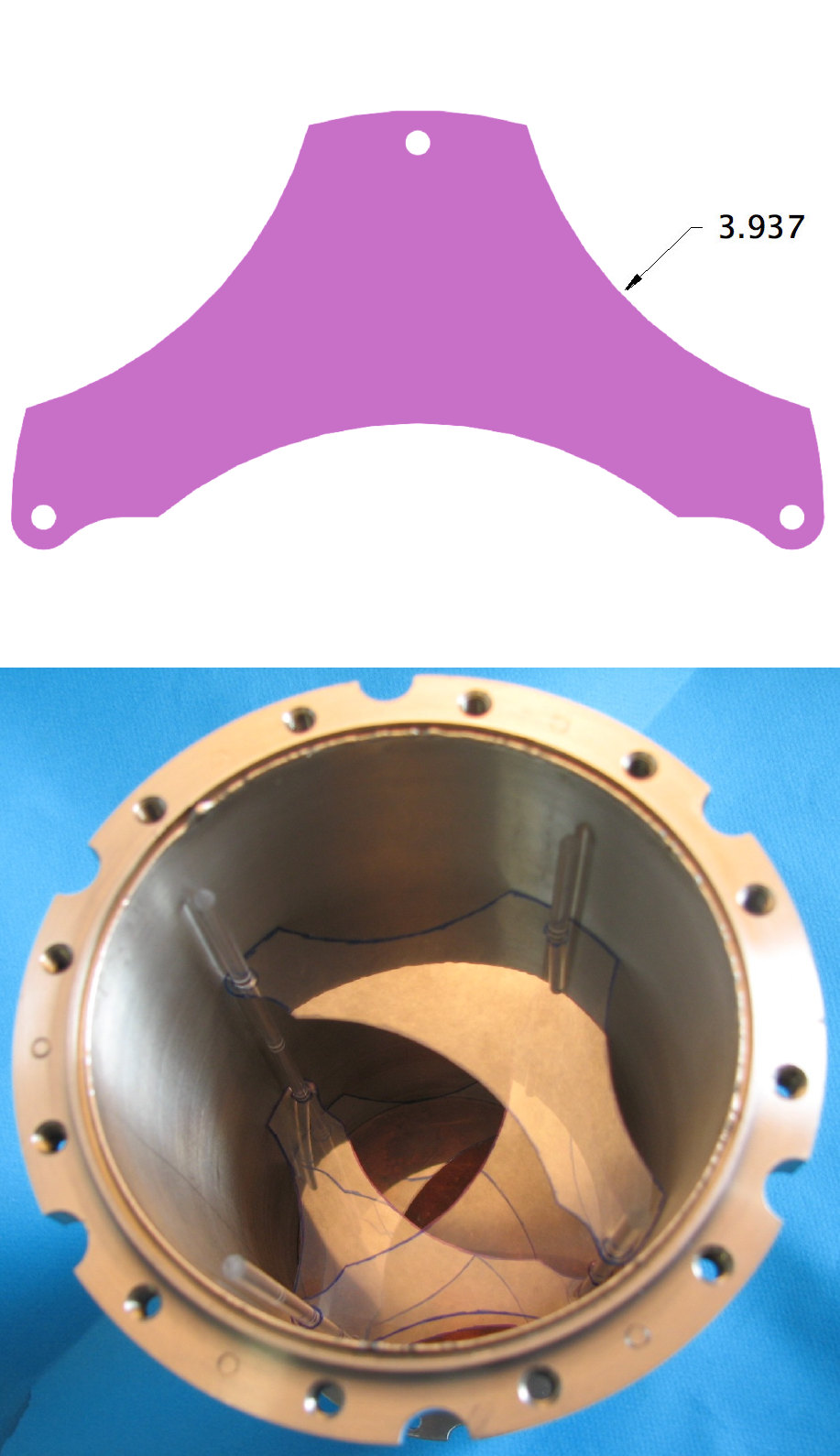}}

\caption{(a) Heat transfer in a cell with four screens breaking the LSC (black symbols,
\textit{Screen cell}) and in a reference cell without screens (green
symbols, \textit{Vintage-cell}). (b) Schematic showing the screen position in the \textit{Screen-cell}. (c) Photograph of the cell interior and schematic of one screen. }
\label{Flo:Fig-LSC-1}
\end{figure}

Is the transition triggered by the shearing of the boundary layers by the LSC ? 
In section \ref{PrSubSec}, we reported some indirect evidence that it was not the case.
We performed here a more direct test of this mechanism by breaking the LSC
with screens to see if the transition was disfavoured. Similar tests have been performed in the past 
to explore the role of the LSC on the hard turbulence regime \cite{ciliberto1996,Xia1997}.
Four croissant-shape
thin horizontal plastic screens are evenly distributed along the
height of the \textit{Screen-cell} (see figure \ref{Flo:Fig-LSC-1}-b/c). The angular distribution of screens
is helicoidal, with a 90 degree angle between consecutive ones.
The surface of each screen equals 33.3\% of the cell cross-section.
Figure \ref{Flo:Fig-LSC-1}-a presents heat transfer measurements in this
cell and in a similar cell without screens but with a residual tilt of 1.3 degree to break the axisymmetry (\textit{Vintage-cell}). The
measurement have been performed for five sets of mean temperature and
mean density conditions: this allows a one-by-one comparison of the experiments
independently of the fluid properties.

Within accuracy, the transition occurs for the same threshold $Ra$ as can be seen for the dataset at $Pr\simeq 0.75$ and $Pr\simeq 0.98$. This suggests that the transition to the Grenoble regime in aspect ratio $\Gamma=0.50$ cells is not triggered by the LSC. This is the main result of this specific experimental study. Additionally, over two decades of $Ra$ above the transition, the heat transfer of both experiments are within 6\%. Thus, a strong alteration of the LSC has a limited effect on heat transfer efficiency in the Grenoble regime in this range of $Ra$ and $Pr$. %

\begin{figure}
\centering
\subfloat[]{\includegraphics[height=0.24\textheight]{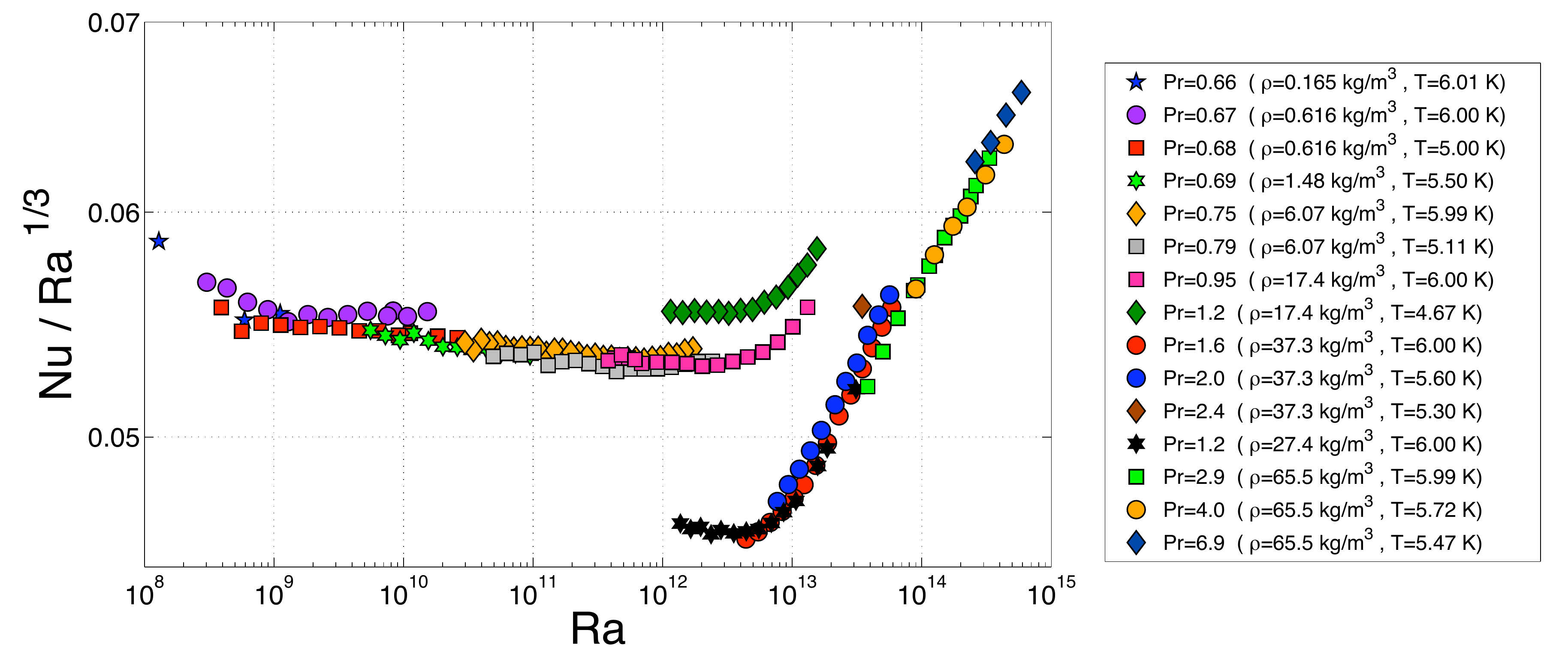}}
\qquad
\subfloat[]{\includegraphics[height=0.24\textheight]{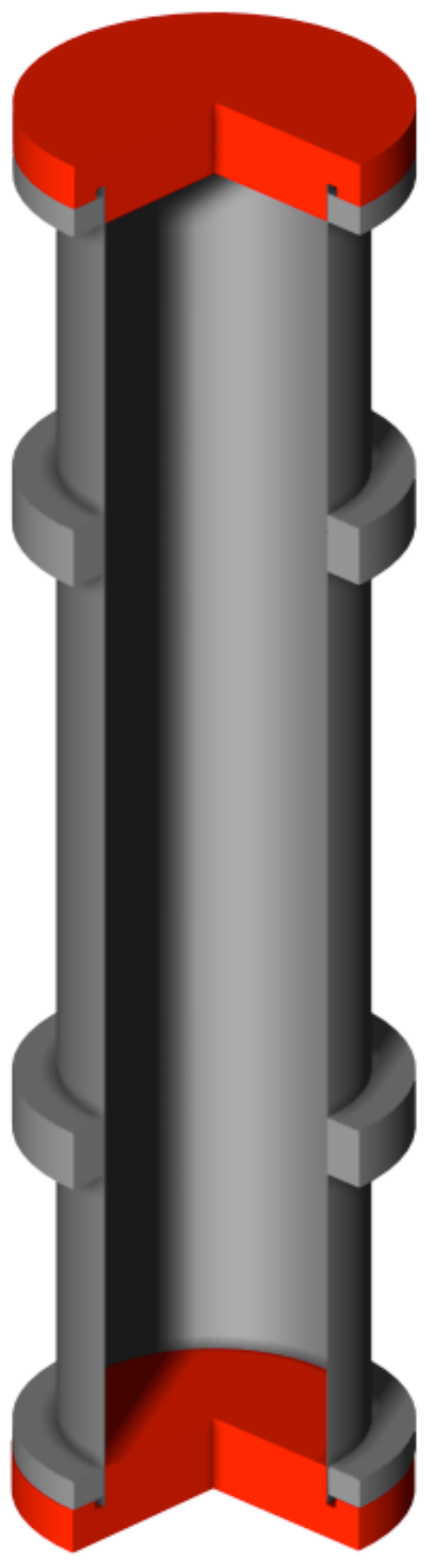}}

\caption{\textit{Cigar-cell} ($\Gamma=0.23$) (a) Compensated heat transfer. The chronological order of the data acquisition is given by the legend, from top to bottom (b) sketch of the cell. }
\label{cigare}
\end{figure}

Finally, a third indication of the limited influence of the LSC on the transition threshold is provided by the $\Gamma =0.23$  \textit{Cigar-cell}. As shown on figure \ref{cigare}, the heat transfer is multi-valued around $Ra\simeq 10^{12}-10^{13}$ with $\sim 14\,\%$ difference in $Nu$ for a given $Ra$. In such an elongated cell, the LSC can be organized in one or several rolls on top of each other. We interpret the bi-valued $Nu$ as a signature of transition from one LSC configuration to another one.  
A similar effect has been reported in the past with few \% difference on $Nu$ in a $\Gamma =0.5$ cell, which was interpreted as the first evidence of the multistability of the LSC in turbulent convection \cite{RocheEPL2002}
 \footnote{Interestingly, although the LSC multistability has been confirmed by simulations \cite{Verzicco2003} and adopted by the community, the original heat transfer measurements have been questioned by some groups failing to reproduce them with water. This may illustrate one of the advantages of cryogenic environment for precise heat transfer measurements: the plates have a very low thermal inertia and a very high diffusivity (compared to \textit{He}) therefore they don't alter the LSC dynamics.}.
The transition to the Grenoble regime happens to occur in the window where $Nu$ is multivalued, giving the opportunity to compare the transition for both configurations of the LSC. Within resolution, the transition occurs for the same $Ra$ and is as steep for the two mean flows.

To conclude this section, we first recall that the heat transfer in the hard turbulence regime is  hardly altered by changes of the LSC. In particular, it well known that a cell tilting or the presence of screens inside the cell motly result in a change of the prefactor of the $Nu(Ra)$ scaling. In this section, we found that this properties is also satisfied in the Grenoble regime.
A second and more surprising conclusion is that the transition to the Grenoble regime is not directly triggered by LSC for $Pr$ of order unity, contrary to what is often assumed in the literature since 2001 when discussing the transition mechanism (e.g. \cite{Chavanne2001,grossmann:PRE2002,Niemela2003,AhlersGrossmannLohse_Review2009}).

\section{\label{sectionSideWall}Grenoble regime and Sidewall}

\subsection{Is the transition altered by lateral confinement?}

\begin{figure}
\centering
\subfloat[]{\includegraphics[height=0.2\textheight]{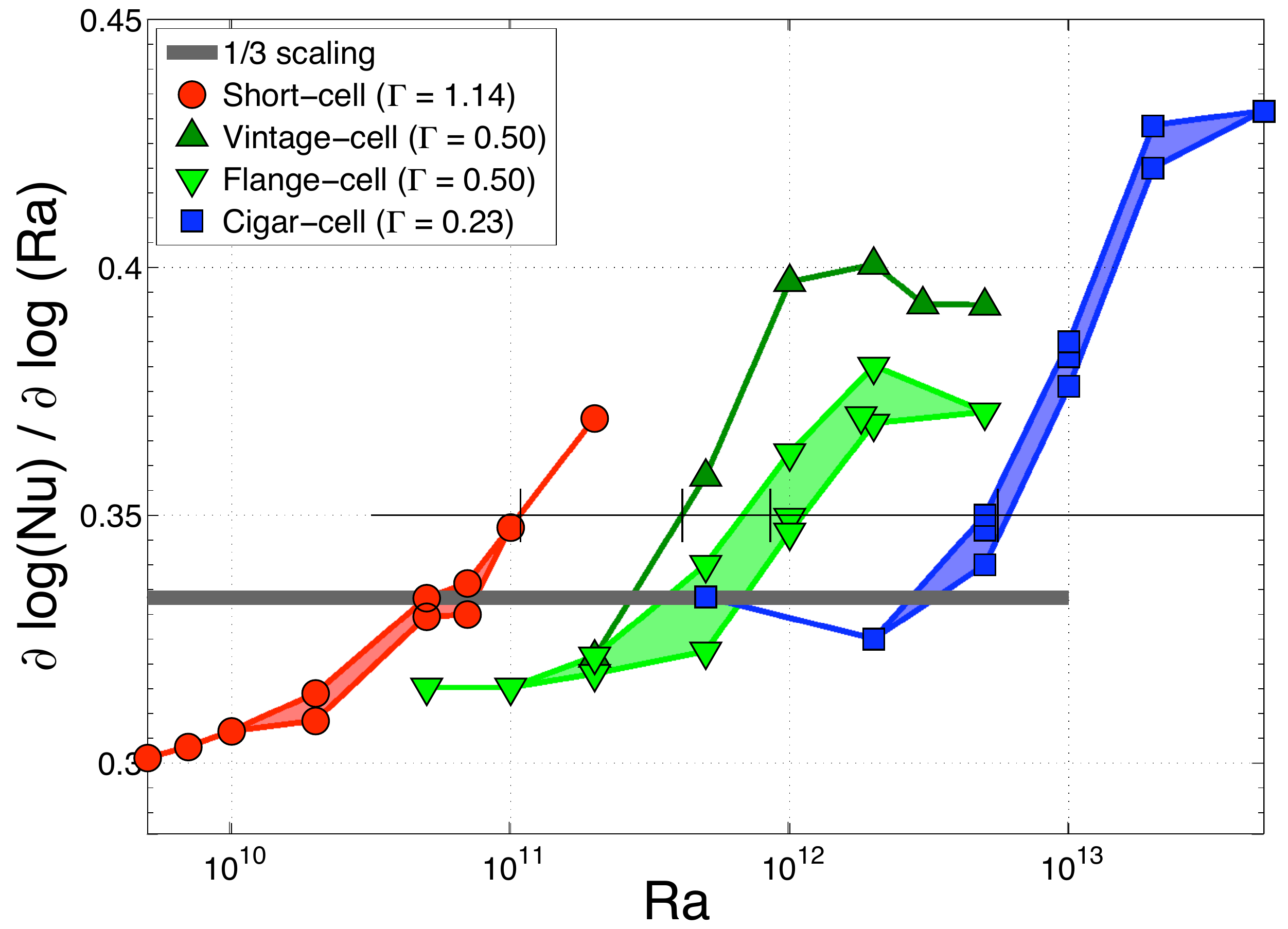}}
\qquad
\subfloat[]{\includegraphics[height=0.2\textheight]{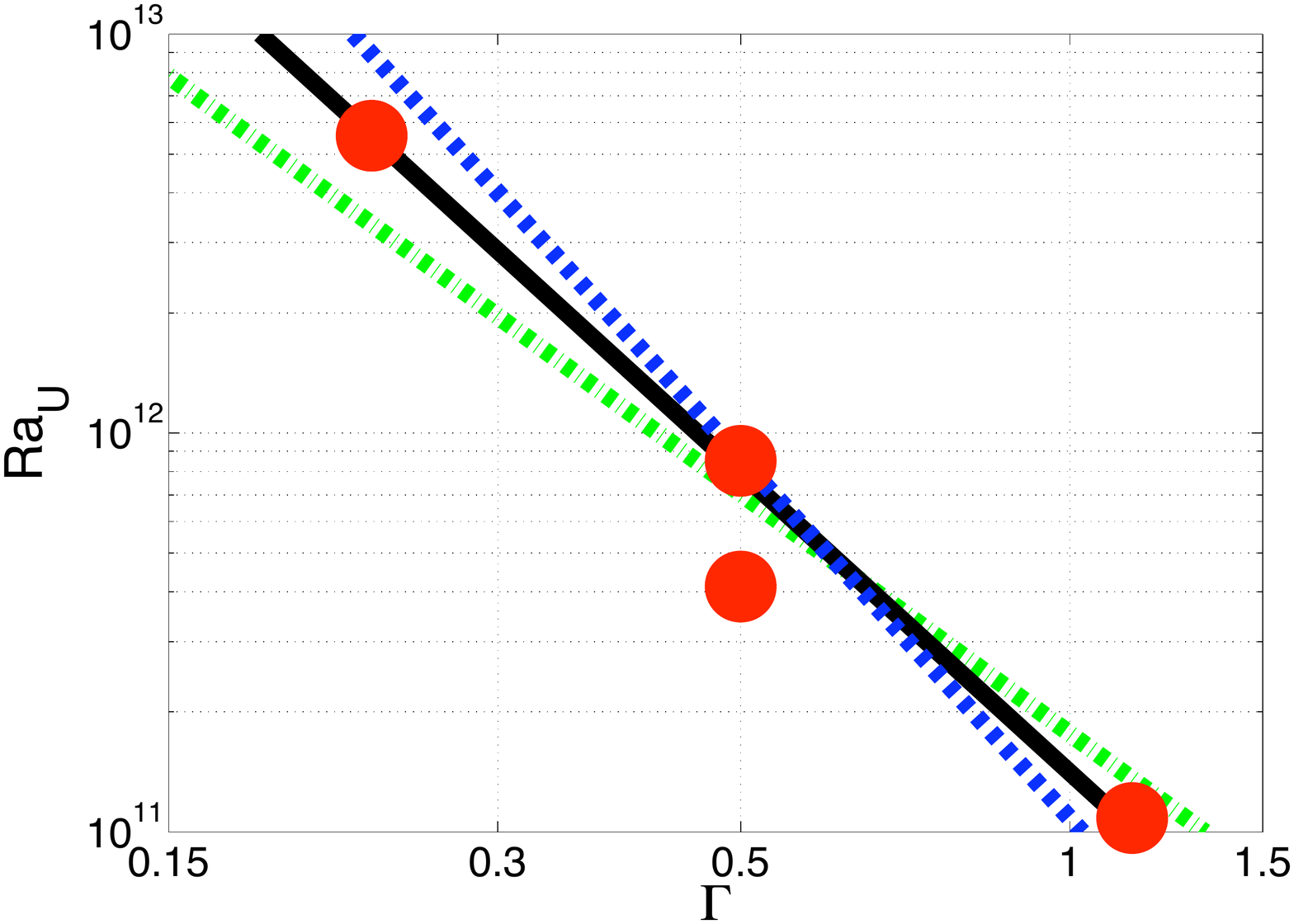}}
\caption{Aspect ratio dependence of the transitional $Ra$. (a) Local scaling exponent  of $Nu(Ra)$ for $Pr=1.5\pm20\%$ in cells with rather similar sidewalls but aspect ratio from $\Gamma=0.23$ to $\Gamma=1.14$  -  (b) $Ra_U$ versus aspect ratio. The lines correspond to $Ra_U\sim \Gamma^{-2}$, $\Gamma^{-2.5}$ and $\Gamma^{-3}$.}
\label{Flo:Fig-Gamma}
\end{figure}

Figure \ref{Flo:Fig-Gamma}-a presents the local scaling exponent of $Nu(Ra)$ in cells with aspect ratios $\Gamma=1.14$ (\textit{Short-cell}), $\Gamma=0.50$ (\textit{Vintage} and \textit{Flange} cells)  and $\Gamma=0.23$ \textit{Cigar-cell}. All these cells have the same diameter $\Phi$, their sidewalls are made of the same material ($\sim0.5\,mm$ thick stainless steel) and the fluid Prandtl number is the same ($Pr=1.5$) within $\pm20\%$. An arbitrary threshold exponent 0.35 is used to define a transition Rayleigh number $Ra_U$ for each cell and this quantity is plotted versus $\Gamma$ on figure \ref{Flo:Fig-Gamma}-b. We find a strong $Ra_U(\Gamma)$ dependence, fittable as $Ra_U\sim \Gamma^{-2.5}$ (solid line) but remains also compatible with a $\Gamma^{-3}$ dependence (blue dash line).
The $Ra_U\sim \Gamma^{-3}$ scaling can be interpreted stating that the transition occurs when a flow length scale that is proportional to the cell height $h$ reaches a constant scale found in the 4 cells. What is this fixed length scale ?

A first hypothesis is that this fixed scale is set by defects of the plates' surface. Systematic roughness and flatness/wavyness characterisations have been performed on all the plates ever operated in Grenoble. We found that the characteristic scales of these defects (see section \ref{sectionManip}) is much smaller than the thermal boundary layer thickness $h/2Nu\sim 200\,\mu m$ at the transition. Moreover, among cells with the same height, we didn't find any correlation between the transition $Ra$ and the plates' roughness although some were 5 times rougher than others. Finally, we note that the Oregon and Trieste cells are made with the same plates : if a plate defect was causing the transition, the Oregon cell should have transited. Rejecting this first hypothesis, we retain the most obvious common length scale shared by the 4 cells: the cells diameter $\Phi$. The $Ra_U\sim \Gamma^{-3}$  scaling can then be reformulated stating that the transition occurs when the Rayleigh number based on $\Phi$ (instead of $h$) reaches a critical value, at least  as first approximation and within the limited range of  aspect ratios $0.23 \le \Gamma \le 1.14$. Surely, this properties associated with the lateral confinement is not expected to extrapolate to aspect ratio much larger than unity.

\subsection{Thermal contribution of the sidewall to the transition.}

We found that confinement by sidewalls disfavours the transition. Is this confinement effect purely geometrical (presence of fixed lateral boundaries) or is it also coupled to the thermal properties of the sidewall ? To explore this possibility, the thermal properties of the sidewall have been altered in three ways :

\begin{enumerate}
\item \textit{Paper-cell}: three  layers of smooth paper has been rolled against a $0.5\,mm$-thick stainless steel sidewall. This results in an increases of the sidewall roughness, a $165\,\mu m$-thick thermal insulation between the flow and the stainless steel sidewall and an extra thermal inertia attached to the sidewall (the inertia of the He trapped in the paper is few tens of times larger than the inertia of the stainless steel).
\item \textit{ThickWall-cell}: the sidewall has been made roughly 4.4 times thicker to increase the sidewall spurious thermal effects
\item \textit{CornerFlow-cell}: a tunable heating (respectively cooling) ring have been installed on the outer side of the sidewall, near the bottom  plate (respectively top plate) connection. This enable to force more or less the corner flows expected at the plate-sidewall angle. An isolated heater was also varnished on the external side of the sidewall, $5\,cm$ above the bottom plate. This heater enables to break the symmetry of the sidewall.
\end{enumerate}

The experiments ii) and iii) have been described in a  conference proceeding \cite{GauthierETC11_2007}. It was found that these alterations of the sidewall have a very limited impact on the transitional $Ra$ and on the strength of the transition (the $Nu(Ra)$ data of the \textit{ThickWall-cell} appears on figure \ref{Flo:Fig-exposants}). The \textit{Paper-cell} experiment is more surprising: as illustrated on figure \ref{Flo:Fig-exposants},   the local exponent of $Nu(Ra)$ increases more slowly with $Ra$ than for the other cells, and the transition seems to occur at slightly higher $Ra$. We have no understanding why the paper layers have such a strong effect on the steepness of the transition, but this results suggests that a thermal interaction between the sidewall and the flow can significantly alter the transition.

As a conclusion on this section, the transition seems significantly disfavoured by the lateral confinement of the sidewall.  A practical consequence is that the transition is not easier to observe in elongated cells, because it occurs at higher $Ra$. We found that the transitional $Ra$ and the transition sharpness can significantly vary from one cell to another depending on the sidewall properties. This observation raises issues on the modelling of sidewall in very high $Ra$ simulations.

\section{\label{discu}Discussion : Grenoble regime Vs. Kraichnan regime}

Over the last 14 years, the Grenoble regime has been characterized in various ways. We summarize the main observed features and discuss them in connection with Kraichnan's prediction.

\begin{enumerate}

\item \textbf{The $Nu(Ra)$ scaling}.

Kraichnan predicted  the following heat transfer law :
\begin{equation}
Nu \sim \frac{Ra^{1/2}}{\left( \log Re \right)^{3/2}}
\label{EqKraichnan}
\end{equation}

We omit a $Pr$-dependent prefactor of the numerator because the $Pr$ dependence of the denominator is not considered in the model. A precise quantitative comparison of (\ref{EqKraichnan}) with the measurements is difficult to justify because this equation  is only valid for asymptotic large $Ra$ (\textit{a-priori}), as Kraichnan insists\footnote{``In view of the inaccuracies inherent in the mixing-length approach, we think that it would be largely illusory to correct discrepancies of this kind by a more careful treatment of the joins between the various asymptotic regions'' (from \cite{Kraichnan1962} p.1386). }. Nevertheless, this equation sets bounds on the local scaling exponent of $Nu(Ra)$ in the region joining the two regimes. Indeed, we expected this exponent to be larger than 1/3 (hard turbulence regime) and lower than 0.5 (asymptotic exponent of (\ref{EqKraichnan})). This window of exponents is consistent with the exponents observed in the Grenoble regime (see figure \ref{Flo:Fig-exposants}).

A more quantitative test of the $Nu(Ra)$ scaling is nevertheless possible. Indeed, in the framework of the Kraichnan model, the boundary layer theory predicts that the denominator in (\ref{EqKraichnan}) should be replaced by a constant if the plates are rough or corrugated. This test was performed (\textit{Corrugated-cell}) and a pure $Nu\sim Ra^{0.5}$ scaling was observed \cite{RochePRE2001}.

\item  \textbf{Evidence of a boundary layer instability}. Analysing the shot noise generated by thermal plumes in the bottom plate, it was found that the Grenoble transition is indeed associated with an instability localized in the boundary layer \cite{GauthierShotNoise:EPL2008}. This result is consistent with the occurrence of Kraichnan's regime, which is characterized by a laminar-to-turbulent transition of the velocity boundary layers laying over the heating and cooling plates. 

\item \textbf{Transition on the fluid's temperature fluctuations}.  Non-invasive measurements within the boundary layers are very delicate due to its small thickness. A direct test of the turbulent state of the boundary layer is therefore difficult. Nevertheless it was found that the temperature fluctuations right above the bottom boundary layer do experience a change of statistics when the transition occurs \cite{GauthierCore:EPL2009} (see also \cite{Chavanne2001}).

\item \label{weakLSC} \textbf{Weak influence of large scale circulation} (LSC). In section \ref{sectionSideWall}, we found that breaking the LSC with screens have a limited impact on the heat transfer, although the LSC must have been significantly altered. Similarly, the transitions to the Grenoble regime had similar characteristics when occurring on two different configurations of the LSC in the \textit{cigar cell}. An increased tilt of the convection cell resulted in a better clamping of the LSC polarisation but did not result in lower $Nu$ as found in the hard turbulence regime.

Finally, in section \ref{sectionPr}, we found that the transition was not occurring for a fixed value of the LSC Reynolds number, at least for $0.98 \le Pr \le 2.9$. The addition of all these observations clearly indicates that the LSC alone is not a transition trigger, as often is assumed in the literature. Once the transition has occurred, we find that the heat transfer is robust to alterations of the LSC, such modifications resulting mostly in new prefactors (of order 1) for the $Nu(Ra)$ law.

\item \textbf{The transition $Ra$}. 
The transition $Ra$ is found typically between $10^{11}$ and $10^{13}$. Cell diameter and sidewall material given, the transition $Ra$ was found to scale like $h^{2.5}$ for  aspect ratios within 0.23 - 1.14. Thus, in this range, the transition Rayleigh number based on the cell diameter (instead of $h$) is nearly constant. This properties of confined cells is not expected to hold for aspect ratio much larger than unity.

In Kraichnan's model, the transition occurs when the eddies located in the bulk of the flow shear the velocity boundary layers beyond their stability point.
As argued in \cite{Chavanne2001}, a transition at $Ra\simeq 10^{11}$ would be compatible with a boundary layer instability that would originate from the shearing by the LSC. %
Following this idea, various estimates based on different hypotheses suggest that any transition occurring in the window $Ra\simeq 10^{11}-10^{15}$ (for $Pr\sim 1$) would also be compatible with Kraichnan's model (e.g. see \cite{Niemela2003, grossmann:PRE2002}).
Unfortunately, we found that the transition cannot be simply triggered by the LSC. These estimations for the transition $Ra$ should therefore be considered with much reserve.
Alternatively, we could speculate that the destabilizing shearing of the boundary layer is caused by the velocity fluctuations above the boundary layers. This scenario is not incompatible with the results of section \ref{sectionLSC}.  The small $Pr$ dependence of the transition near $Pr\simeq 2$ (section \ref{PrSubSec}) could result from the proximity with the low $Pr$ region. Thus, strictly speaking, the present results are not ruling out Kraichnan's model,  but they are not supporting it either.

As a final remark possibly related to the Grenoble regime, the thermal and velocity boundary layers have been characterized at intermediate $Ra$ (where it is thick enough to be resolved) in room temperature experiments. Extrapolation of their properties over three decades of $Ra$ suggests the occurrence of a transition compatible with Kraichnan's views near $Ra\simeq10^{13}$ \cite{Belmonte:PRL1993,Sun:JFM2008}.

\item \textbf{Transition in the $Ra-Pr$ parameter space}.  In section \ref{sectionPr}, we showed that all $\Gamma=0.5$ experiments reporting a transition are localized in the same region of the $Ra-Pr$ parameters space  and that nearly all the experiments without transition fall outside this region. There is no understanding of this parameter space. Updating Kraichnan's model using the knowledge gained over the last 50 years would certainly be highly interesting.

\item \textbf{A threshold exponent 1/3 ?} For most Grenoble cells, the transition seems to occur abruptly when the local exponent of $Nu(Ra)$ reaches the value $1/3$, as illustrated in figure \ref{Flo:Fig-exposants}.  Reversely, the exponent $1/3$ is not reached  in most of the very high $Ra$ experiments which don't report a transition, as illustrated in figure \ref{Flo:Fig-NuRa-compil}. Interestingly, the exponent $1/3$ is found in the Malkus model of uncoupled boundary layers \cite{Malkus:1954heatTransport}. We have no interpretation for these coincidences.

\item \textbf{Thermal interaction with the sidewall}. Finally, we recall that the insertion of paper layers between the stainless steel sidewall and the fluid results in a significantly less sharp transition. To the contrary, two other changes of the sidewall thermal properties (using a thicker wall and modifying the temperature distribution along the wall) only had a limited effect. This result is not understood but reveals that  the sidewall plays some role in the transition, as already suggested by the comparison of cells with different aspect ratios.

\end{enumerate}

\section{Conclusion}

We presented a systematic study of the convection regime reported in Grenoble in 1996, and then named \textit{the Ultimate regime} of convection. In particular, we characterized the conditions for the triggering of this regime. Among the results, we showed that  all Rayleigh-Bénard experiments conducted at very high $Ra$ using cryogenic helium are consistent if we assume that low $Pr$ tend to disfavour the transition in aspect ratio $\Gamma=0.5$ cells. We also found that the large scale circulation present in the cells does not play a key-role in triggering the transition, contrary to a common assumption. Reversely, we found that the sidewall has an unexpected effect on this transition.

Several evidences suggest that Grenoble's regime corresponds to Kraichnan's prediction and no experimental fact seems incompatible with such an interpretation. Nevertheless, the conditions for the triggering of this regime are obscure and sometimes surprising. Besides, a few experimental facts cannot be directly explained using the genuine Kraichnan model. Further experimental investigations are clearly needed.

On the theoretical side, the Kraichnan regime is the only elaborated model available to interpret the Grenoble regime. Alternative scenarios of boundary layer instability probably deserve to be explored, aside Kraichnan's paradigm or in a complementary fashion. We hope that the results presented in this work will set useful bounds for such alternative models.

\ack
We thank B. Castaing, F. Chillà, R. Du Puits, B. Hébral and B. Chabaud for scientific discussions and support. We acknowledge the laboratory cryogenic staff, and in particular G. Garde, for providing us with most of the convection cells presented in this work. We don't forget V. Arp who kindly shared us with some helium properties fits, and the students R. Bois, J. Muzellier and E. Waites who contributed at some point to these investigations.

\clearpage

\appendix

\section{\label{nob}Grenoble regime and the Boussinesq's approximation}

We report here two systematic and complementary studies of non-Boussinesq deviations at very high $Ra$. An experimental study explores the validity of the ``constant fluid properties" approximation \cite{Oberbeck1879}, while the theoretical study addresses the validity of all the other approximations required  to obtain Boussinesq's set of equations \cite{Boussinesq1903}. The ``constant fluid properties'' approximation in the context of very high $Ra$ convection has already been discussed in the literature (e.g. \cite{Niemela:2006,Chavanne2006}).

\subsection{The variation of fluid properties}

In given experimental conditions, the variations of the different fluid properties across the cell occur simultaneously. It is therefore convenient to parametrize them with single parameter. A convenient 
one is the first-order-approximation of the density variation, $\delta\rho/\rho\simeq\alpha\Delta$.
A more stringent parameter $\Upsilon$ is the maximum deviation among
the five parameters coming into the definition of $Ra$, $Pr$ and $Nu$: the density $\rho$, the thermal expansion $\alpha$, the molecular
conductivity $k$, the isobaric heat capacity $c_{p}$ and the viscosity
$\eta$. In practice, we defined it as $\Upsilon=2\cdot\max(|\delta A_{top}|/A_{top},|\delta A_{bottom}|/A_{bottom})$
where $A\in\left\{ \rho,\alpha,k,c_{p},\eta\right\} $ and where the
indices $top$ and $bottom$ represents the top and bottom boundary
layers. Using these two non-Boussinesq parameters, we now compare the
datasets of the \textit{Flange-cell}, Trieste $\Gamma=1$ and Oregon experiments. Among our cells, the \textit{Flange-cell} is chosen because it benefited of the most extensively exploration of the parameter space.
All three cells used the same fluid and therefore they are expected to experience
somehow similar variation of fluid properties. We recall that the Oregon
cell doesn't transit at very high $Ra$ contrary to the two others.

\begin{figure}
\center
\includegraphics[width=.6\textwidth]{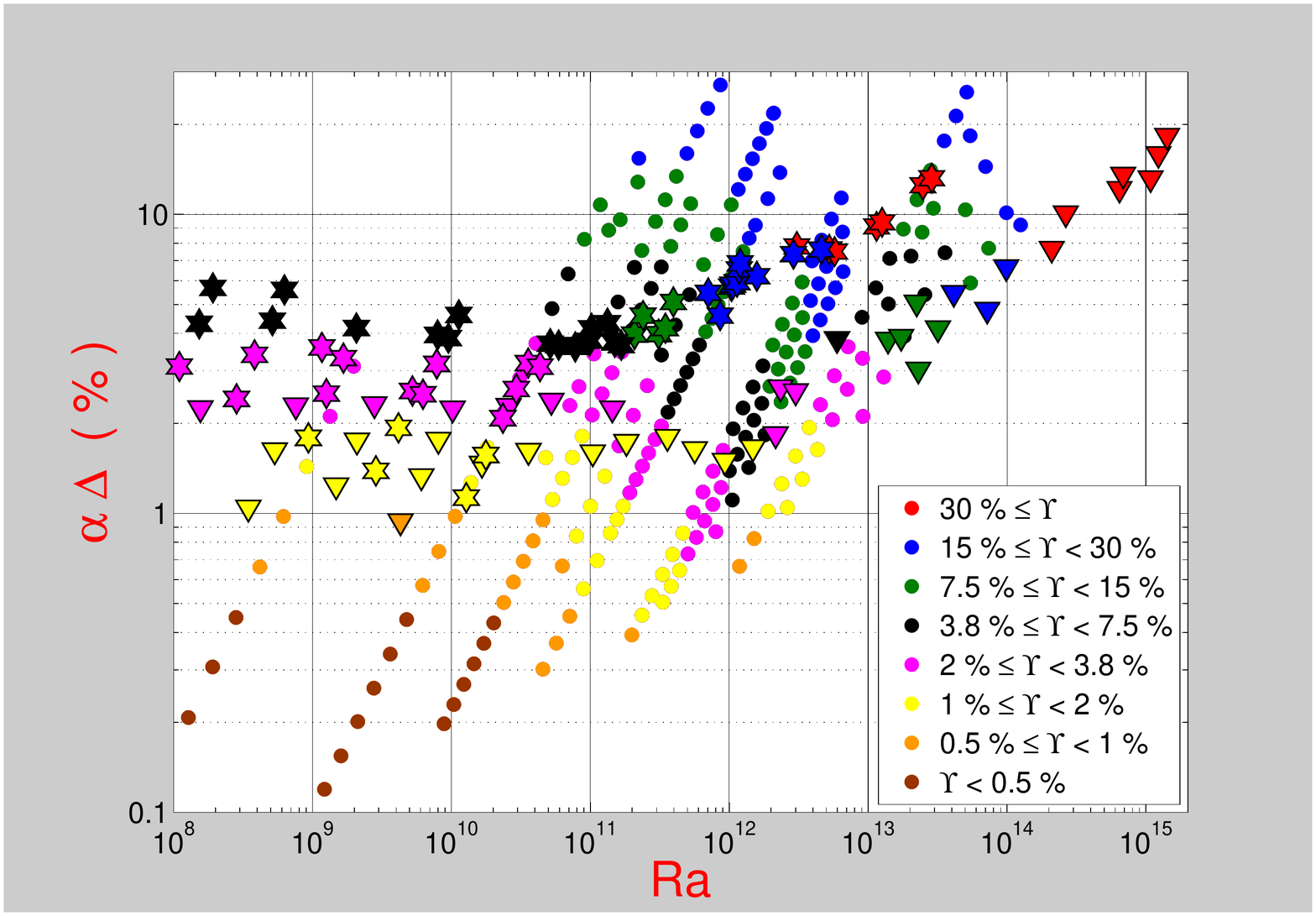}
\caption{Non-Boussinesq variation of fluid properties in the \textit{Flange-cell} (disks), Oregon (triangles) and Trieste $\Gamma=1$ (stars) experiments. The non-Boussinesq parameter  is defined in the text.}
\label{Flo:Fig-Boussi-1}
\end{figure}

Figure \ref{Flo:Fig-Boussi-1} shows the non-Boussinesq parameters $\alpha\Delta$
(y-axis) and $\Upsilon$ (color code) for the three experiments.
At first, for any given $Ra$,
the non-Boussinesq deviations in Grenoble 20-cm-high cell can be significantly
smaller than the corresponding deviations in Trieste 50-cm-high cell,
and at least as good as the 100-cm-high cell of Oregon. This contradicts
a widespread idea that larger cells are necessarily ``more'' Boussinesq
for a given $Ra$.

More interesting, for a given $Ra$, the non-Boussinesq parameter
$\alpha\Delta$ of the Grenoble  experiment is varied over
up to 1.5 decade, reaching values equal or above the ones reached
in the Oregon cell. At $Ra=2\cdot10^{12}$ (for example), the non-Boussinesq
parameters $\alpha\Delta$ and $\Upsilon$ are close to 2\% in the
Oregon experiment. At the same $Ra$, these parameters are spanning
a range from 1\% up to nearly 30\% in Grenoble's cell.
Over all this range of non-Boussinesq parameters, Grenoble Ultimate regime
is present in Grenoble experiment but not in the Oregon one. 
This shows that the non-Boussinesq deviations associated
to fluid properties variations cannot cause the qualitative difference
between these two experiments.

\begin{figure}
\centering
\subfloat[]{\includegraphics[height=0.2\textheight]{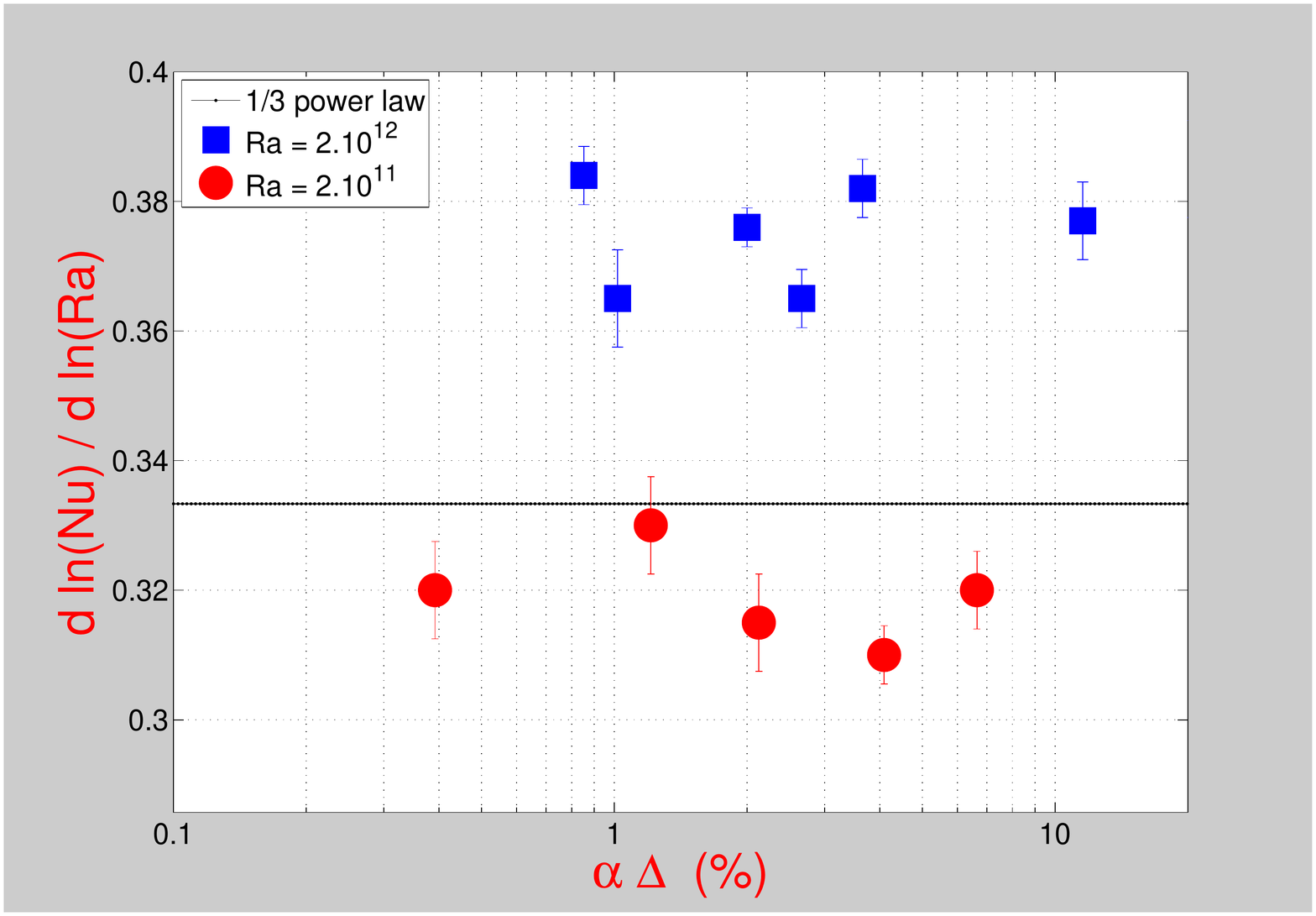}}
\qquad
\subfloat[]{\includegraphics[height=0.2\textheight]{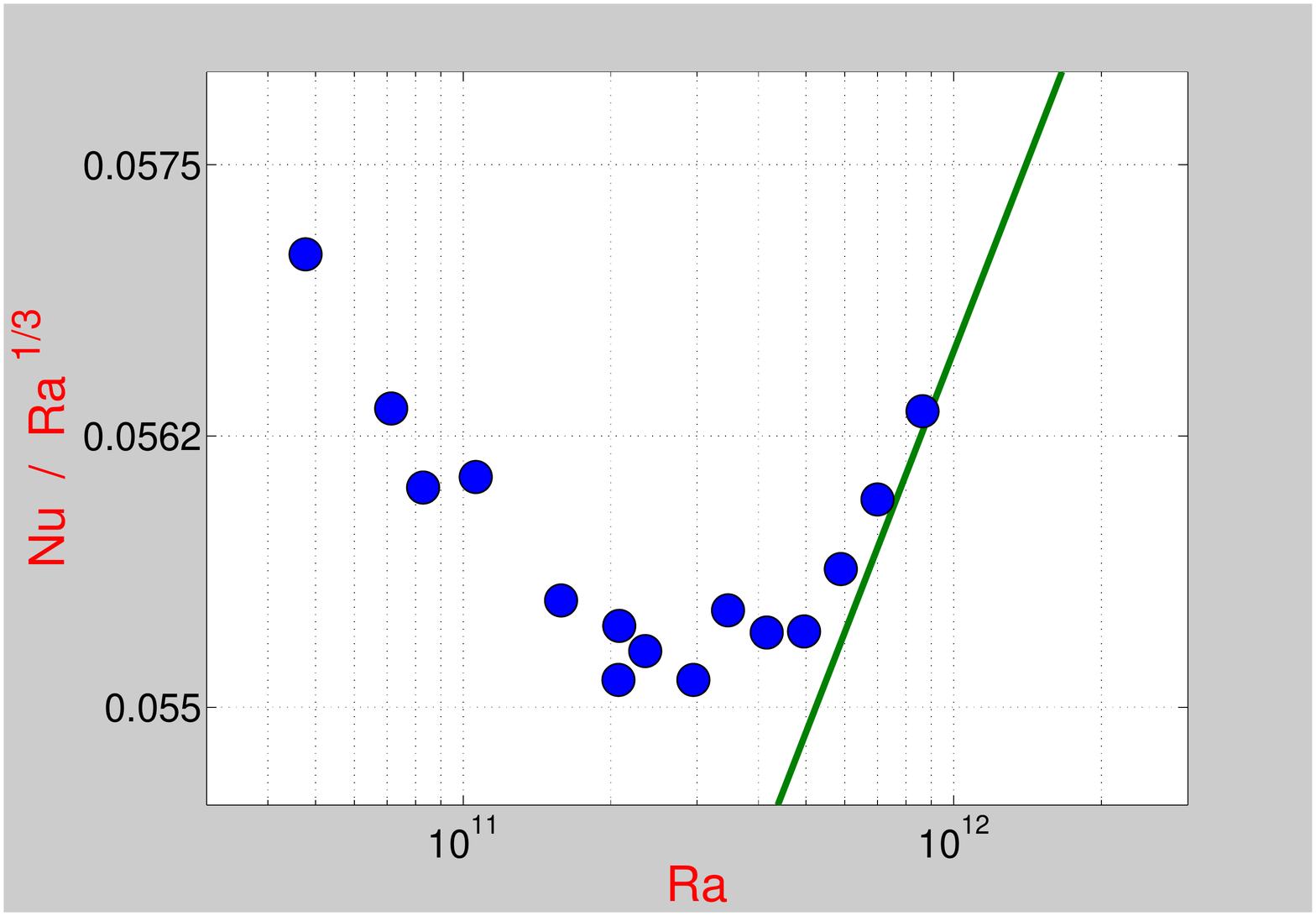}}
\caption{(a) Local exponent of the $Nu(Ra)$ law versus the non-Boussinesq
parameter $\alpha\Delta$. The Rayleigh numbers $Ra=2\cdot10^{11}$ and
$2\cdot10^{12}$ were chosen on each side of the transition to the Ultimate
regime. (b) Evidence of the transition at a mean density of 18.69 $kg/m^3$ and mean temperature of 11.4 K, that is at about a quarter of the critical density and at twice the critical temperature.}
\label{FigBoussi23}
\end{figure}

Figure \ref{FigBoussi23}(a) shows the local scaling exponent of $Nu(Ra)$
in Grenoble cell for two $Ra$ : below and  above the transition. Each exponent was fitted manually using
the subsets of ($Ra$, $Nu$) data obtained at constant mean density and
mean temperature. %
 For these two specific $Ra$, the non-Boussinesq parameter $\alpha\Delta$ varies
over more than one decade without causing any noticeable change of scaling exponent. It is worth pointing out that the Grenoble regime is clearly
evidenced with a non-Boussinesq parameter $\alpha\Delta$ smaller than 1\%.

\subsection{Non-Boussinesq deviations without fluid properties variation}

Boussinesq's approximation doesn't only consist in neglecting variation
of fluid properties and modelling at first order the buoyant term.
It also requires a decoupling of the heat
balance equation with the flow mechanical energy. For example, the
heating produced by viscous dissipation and temperature fluctuations
induced by pressure fluctuations are neglected in this approximation.
In his famous book \cite{Tritton1988}, D. J. Tritton discusses in
details the applicability of this approximation at low $Ra$, where
gradient and time derivatives can be estimated using integral length
and time scales $h$ and $h^{2}/\nu Re$. In the turbulent regimes,
these gradients could a-priori become significantly larger due the
smaller characteristic scales and Tritton's criteria are no longer
useful. Using the present knowledge of scaling laws in turbulent convection,
we derived a set of criteria for the applicability of Boussinesq's
approximation in a fluid with constant properties and $Pr$
of order $1-10$. Details of the derivation are given in \cite{RocheBoussinesq:2007}. The result 
is a set of four criteria :

\begin{center}
$\left\{\begin{array}{lcll}
\alpha\Delta &&\qquad& \ll1\\
\alpha\Delta_{h}&\times&\case{PrRe^{2}}{Ra\theta_{rms}^{\star}} & \ll1\\
\case{\Delta_{h}}{T_{0}}&\times&\left(\case{Pr^{2}Re^{3}}{Ra\theta_{\partial t}^{\star}}+\case{PrRe}{4Nu^{2}}\right) & \ll1\\
\case{\Delta_{h}^{2}}{T_{0}\Delta}&\times&\case{PrRe}{2\theta_{\partial t}^{\star}} & \ll1 \end{array}\right.$
\par\end{center}

\noindent where $\theta_{rms}^{\star}$ is a dimensionless estimation of
the temperature fluctuations in the bulk of the flow and $\theta_{\partial t}^{\star}$
is an estimation of its associated time derivative $\partial\theta/\partial t(h^{2}/\Delta\kappa)$.
The adiabatic gradient $\Delta_{h}/h,$ the compressibility $\chi$, the heat capacities ratio $\gamma=c_{p}/c_{v}$ and the mean temperature
$T_{0}$ are related by :

\begin{center}
$\alpha\Delta_{h}=\alpha^{2}gT_{0}h/c_{p}=\rho_{0}g\chi h(1-\gamma^{-1})\sim\rho_{0}g\chi h$
\par\end{center}

Physically, the first criterion is required for incompressibility
in the boundary layer and it was considered in the first part of this appendix.
The second criterion allows to
neglect variations of density due to the pressure variations, versus
those due to temperature variations. It also allows to neglect the
cooling/heating associated with the pressure variations experienced
by a fluid particle in the heat transport equation. The two other criteria,
indirectly associated with the stratification in the cell, result
from various physical contributions. Using fits for $Re$, $Nu$, $\theta_{rms}^{\star}$
and $\theta_{\partial t}^{\star}$, the four criteria above become:

\begin{center}
$\left\{\begin{array}{lcll}
\alpha\Delta &&\qquad& \ll1\\
\left(\alpha\Delta_{h}\right)&\times&0.1Ra^{0.13} & \ll1\\
\left(\case{\Delta_{h}}{T_{0}}\right)&\times&\left[\case{Ra^{0.1}Pr^{-0.25}x}{200}+10Ra^{-0.17}Pr^{0.25}\right] & \ll1\\
\left(\case{\Delta_{h}^{2}}{T_{0}\Delta}\right)&\times&0.1Ra^{0.1}Pr^{0.25} & \ll1\end{array}\right.$
\par\end{center}

Our main interest is to see if a violation of the criteria \#2, \#3
or \#4 is correlated with the occurence/inhibition of the transition
to the Grenoble regime. We find that
up to $Ra\simeq10^{16}$, the Oregon and Grenoble \textit{He} experiments fullfil
the last three criteria provided that the first one is fulfilled \cite{RocheBoussinesq:2007}  
\footnote{Above $Ra\simeq10^{16}$, the criteria \#2 is violated in the Oregon
experiment. A side consequence is that increasing the height of \textit{He}
cell beyond $1\,m$ will only allow to increase the maximum $Ra$ (within Boussinesq
conditions) like $h^{2}$ and not $h^{3}$.%
}. As a conclusion, the type of non-Boussinesq deviation considered in
this subsection cannot explain the puzzle at very high-$Ra$.

\subsection{A critical point effect ?}

The convection literature repeatily states that the cryogenic \textit{He}
experiment are performed ``close to the critical point'' and suggest
that some unknown critical point artefacts could somehow alter heat
transfer measurements at very high $Ra$.

Firstly, we note that no precise statement along this line has ever been published.

Secondly, this hypothesis has been discussed in  \cite{Chavanne2006} comparing the datasets of Chavanne \etal \cite{Chavanne1997}, Oregon \cite{Niemela2000} and Chicago \cite{WuTHESE}. The conclusion was that proximity to the liquid-vapor coexistence curve cannot cause the difference in heat transfer.

A third argument can be raised using the \textit{Flange-cell} datasets. The ``distance'' to the critical point ($T_{c},$$\rho_{c}$)
can be assessed quantitatively using the reduced temperature $\mid T-T_{c}\mid/T_{c}$ and reduced density $\mid\rho-\rho_{c}\mid/\rho_{c}.$ It is well known that critical point divergence phenomena
becomes significant when both reduced parameters are significantly
smaller than unity. In \cite{Ashkenazi1999} for example, the reduced temperature is made smaller than $10^{-2}$ to experience compressibility effect. On the other hand, the reduced
temperature and density of water in traditional convection experiments
are about 0.5 and 2, and one can safely consider these experiments as far away from the critical point.
For $Ra=10^{13}$, the reduced density in the
Oregon cell is about 0.9 (that is $\rho\simeq\rho_{c}/10$): undoubtly
a critical point artefact cannot explain the absence of transition
in this cell. In the Grenoble \textit{Flange} cell, the transition was evidenced
for reduced temperatures up to 1.2 ($T\simeq2.2\, T_{c}$) and a reduced
density close to 0.7 ($\rho\simeq\rho_{c}/3.7$), as shown on figure
\ref{FigBoussi23}(b). The occurrence of the transition cannot be
seriously attributed to a critical point artefact.

As a conclusion, it seems very unlikely that any sort of non-Boussinesq deviation could explain the apparent scatter of heat transfer measurements at very high $Ra$.

\end{document}